\documentclass[12pt]{article}

\usepackage{scicite}

\usepackage{times}

\usepackage{graphicx}

\newcommand{\aap}{Astron.\ Astrophys.}
\newcommand{\aj}{Astron.\ J.}

\newcommand{\araa}{Ann.\ Rev.\ Astron.\ Astrophys.}
\newcommand{\mnras}{Mon.\ Not.\ Roy.\ Astron.\ Soc.}

\newcommand{\pra}{Phys.\ Rev.\ A}

\newcommand{\prl}{Phys.\ Rev.\ Lett.}

\usepackage[usenames,dvipsnames]{color}

\usepackage{CJK}

\newcommand{\lsim}{\,\rlap{\raise 0.35ex\hbox{$<$}}{\lower 0.7ex\hbox{$\sim$}}\,}
\newcommand{\gsim}{\,\rlap{\raise 0.35ex\hbox{$>$}}{\lower 0.7ex\hbox{$\sim$}}\,}

\newenvironment{sciabstract}{%
\begin{quote} \bf}
{\end{quote}}

\title{A limit on variations in the fine-structure constant from spectra of nearby Sun-like stars}

\author
{Michael T. Murphy,$^{1\ast}$ Daniel A. Berke,$^{1}$ Fan Liu (刘凡),$^{1}$ Chris Flynn,$^{1,2}$ \\
Christian Lehmann,$^{1}$ Vladimir A. Dzuba,$^{3}$ Victor V. Flambaum$^{3}$\\
\\
\normalsize{$^{1}$Centre for Astrophysics and Supercomputing, Swinburne University of Technology,}\\
\normalsize{Hawthorn, Victoria 3122, Australia}\\
\normalsize{$^{2}$ARC Centre of Excellence for Gravitational Wave Discovery,}\\
\normalsize{Hawthorn, Victoria 3122, Australia}\\
\normalsize{$^{3}$School of Physics, University of New South Wales, Sydney,}\\
\normalsize{NSW 2052, Australia}\\
\\
\normalsize{$^\ast$To whom correspondence should be addressed; E-mail:  mmurphy@swin.edu.au.}
}

\date{}

\begin{document} 

\begin{CJK*}{UTF8}{gbsn}


\baselineskip24pt


\maketitle


\begin{sciabstract}
  The fine structure constant, \boldmath{$\alpha$},
  sets the strength of the electromagnetic force. The Standard Model
  of particle physics provides no explanation for its value, which
  could potentially vary. The wavelengths of stellar absorption lines
  depend on \boldmath{$\alpha$}, but are subject to systematic effects
  owing to astrophysical processes in stellar atmospheres. We measured
  precise line wavelengths using 17 stars, selected to have almost
  identical atmospheric properties to those of the Sun (solar twins), which
  reduces those systematic effects. We found that \boldmath{$\alpha$}
  varies by \boldmath{$\lsim$}50 parts-per-billion (ppb) within 50
  parsecs from Earth. Combining the results from all 17 stars provides
  an empirical, local reference for stellar measurements of
  \boldmath{$\alpha$} with an ensemble precision of 12\,ppb.
\end{sciabstract}



\clearpage

The Standard Model of particle physics contains parameters known as
``fundamental constants''. These include the coupling strengths of the
known physical forces; the strength of electromagnetism is set by the
fine-structure constant, $\alpha\equiv e^2/\hbar c$ where $e$ is the
elementary charge, $\hbar$ is the reduced Planck constant and $c$ is
the (vacuum) speed of light. These dimensionless numbers are referred
to as ``fundamental'' because the Standard Model does not predict
their values. They are usually assumed to be universal constants,
i.e.\ they do not depend on other (unknown) physics. Their values can
only be established experimentally, and testing their constancy
requires measurements under a wide range of physical conditions, such
as different times, distances, gravitational potentials etc..
Measurements of laboratory atomic clocks have set an upper limit on
relative variations in $\alpha$ to $\lsim 10^{-18}$\,yr$^{-1}$ over
several years \cite{Lange_2021PhRvL.126a1102L}. On cosmological time
and distance scales, absorption lines of distant gas clouds in the
spectra of background quasars limit relative variations in $\alpha$ to
$\lsim 1$ parts per million (ppm)
\cite{Kotus_2017MNRAS.464.3679K,Murphy_2017MNRAS.471.4930M,Murphy_2022A&A...658A.123M}. A study of giant stars within the Milky Way has set similar limits of $\lsim 2$--6\,ppm
\cite{Hees_2020PhRvL.124h1101H,Hu_2021MNRAS.500.1466H}.

Any variation in $\alpha$ would alter the energy levels of atoms and
ions in characteristic ways \cite{Dzuba_1999PhRvL..82..888D}. The rest-frame wavenumber of an absorption or emission line, $\omega_{\rm obs}$, would shift from its 
laboratory value, $\omega_0$, in proportion to the relative change
$\Delta\alpha/\alpha \equiv (\alpha_{\rm obs}-\alpha_0)/\alpha_0$:
\begin{equation}\label{e:daa}
\frac{\Delta v}{c} \equiv \frac{\omega_0-\omega_{\rm obs}}{\omega_0} \approx -2 \frac{\Delta\alpha}{\alpha} Q\,,
\end{equation}
where $\alpha_0$ and $\alpha_{\rm obs}$ are the laboratory and
observed values of $\alpha$, respectively, $\Delta v$ is the line
shift in velocity units, and the sensitivity coefficient, $Q$,
describes how much a given line shifts to the blue (for positive $Q$)
or red. The approximation is valid for $\Delta\alpha/\alpha \ll 1$. In
practice, the velocity shifts are measured for multiple lines and
different atoms and ions, which is known as the Many Multiplet
method. Utilising lines with a wide variety of $Q$ coefficients
increases the sensitivity to variations in $\alpha$.

Sun-like stars are a potentially suitable target for the Many
Multiplet method: their spectra contain thousands of narrow,
well-defined, strong (but unsaturated) absorption lines (Fig.\ 1A).
The observed wavelengths of these lines could, in principle, be
compared to their laboratory values while simultaneously accounting
for the star's radial velocity. However, this simple approach is
limited by large systematic errors: several physical mechanisms can
shift the lines by up to $\sim$700\,m\,s$^{-1}$ from their laboratory
wavelengths, and the line profiles are asymmetric because they arise
over a range of depths in stellar atmospheres
\cite{Dravins_1982ARA&A..20...61D,GonzalezHernandez_2020A&A...643A.146G}.
These effects produce velocity shifts between lines, typically
$\Delta v \sim 250$\,m\,s$^{-1}$ \cite{Dravins_1982ARA&A..20...61D},
which is equivalent to $\Delta\alpha/\alpha \sim 6$\,ppm for a typical
range in $Q$ coefficients of $\approx$0.07
\cite{Dzuba_2022PhRvA.105f2809D}.  Direct comparison of absorption
lines in a single giant star to laboratory values has already reached
this systematic error limit \cite{Hees_2020PhRvL.124h1101H}.

We adopt an alternative technique that compares absorption lines
between stars that have intrinsically similar spectra, thereby
eliminating the need to compare with laboratory wavelengths. The
atmospheric spectrum of an isolated main-sequence star depends
primarily on its mass and heavy-element content, which determine three
primary observable parameters: the effective temperature
$T_{\rm eff}$, iron metallicity [Fe/H], and surface gravity $\log g$.
We restrict our analysis to solar twins -- defined as stars with these
parameters within 100\,K, 0.1\,dex and 0.2\,dex of the Sun's values,
respectively. Spectra of two solar twins used in our analysis are
shown in Fig.\ 1A. We measure the velocity-space separations of pairs
of lines, then compare the same sets of lines between stars (Fig.\
1B). This approach reduces the systematic errors from astrophysical
line shifts and asymmetries because of the similarity of their stellar
parameters. The use of pairs of lines removes any dependence on the
stars' radial velocities, including any variations that could be
caused by an orbiting companion (e.g.\ in a planetary or binary
stellar system).  For main-sequence stars, line shifts and asymmetries
are observed to be correlated with the line's optical depth and
wavelength \cite{Dravins_1982ARA&A..20...61D}, so we select pairs with
similar absorption depths (within 20\%) and small separations
($<$800\,km\,s$^{-1}$, equivalent to $\approx$13\,\AA\ at 5000\,\AA)
\cite{Berke_2022a,Berke_2022b,materials}. We chose these values to
reduce the systematic effects while maintaining sensitivity to
variations in $\alpha$ between stars.

We applied this solar twins method to archival solar twin spectra from
the High Accuracy Radial velocity Planet Searcher (HARPS) spectrograph
mounted on the European Southern Observatory (ESO) 3.6\,m telescope at
La Silla Observatory, Chile. HARPS is highly stable over time
\cite{Mayor_2003Msngr.114...20M} and its wavelength scale has been
precisely characterised by using laser frequency combs
\cite{Wilken_2010MNRAS.405L..16W,Milakovic_2020MNRAS.493.3997M}. This
sets an instrumental systematic error limit of $\sim$2--3\,m\,s$^{-1}$
in the velocity separations of line pairs \cite{Berke_2022b}. To reach
this level we restricted our analysis to HARPS exposures corrected for
non-uniform detector pixel sizes (corrections $\sim$25\,m\,s$^{-1}$)
\cite{Coffinet_2019A&A...629A..27C,materials}, and applied a further
correction for sparsely-sampled wavelength calibration (corrections
$\sim$5\,m\,s$^{-1}$) \cite{Milakovic_2020MNRAS.493.3997M,materials}.

We selected 16 bright (i.e.\ nearby) solar twins with HARPS spectra,
with signal-to-noise $\textrm{SNR} > 200$ per 0.8\,km\,s$^{-1}$ pixel,
plus the Sun via its reflection from the asteroid Vesta (with
$\textrm{SNR} > 150$) (Table S1) \cite{materials}. With these SNRs,
the statistical uncertainty in the velocity separation of two
unresolved absorption lines is $\sim$25\,m\,s$^{-1}$ from a single
exposure, assuming that they absorb 50\% of the stellar flux at their
cores \cite{Brault_1987AcMik...3..215B}. The median number of HARPS
exposures available was 10 exposures per star (range of 1--138), so by
combining results from multiple exposures we expect median statistical
uncertainties to reduce to $\sim$8\,m\,s$^{-1}$ per line pair, per
star. By averaging over the sample of 17 stars, the uncertainty
approaches that imposed by the available instrument calibration
\cite{Berke_2022b}.

From 8843 lines listed in a solar atlas
\cite{Laverick_2018A&A...612A..60L}, we select 22 that are separated
from each other and not blended with other nearby stellar or telluric
(Earth atmosphere) lines \cite{Berke_2022b,materials}. All 22 are
strong but unsaturated, absorbing 15--90\% of the continuum in the
HARPS spectrum of the Sun. The 22 lines, which form 17 different pairs
(some of which share common lines), arise from neutral atoms Na, Ca,
Ti, V, Cr, Fe and Ni, plus singly-ionised Ti. Their $Q$ coefficients
have been calculated previously \cite{Dzuba_2022PhRvA.105f2809D}. The
17 pairs of lines have a wide range of sensitivity to $\alpha$
variation, with differences in $Q$ within each pair from $-0.08$ to
$+0.18$ \cite{Dzuba_2022PhRvA.105f2809D}.

Pair separations were measured using a fully automated process for all
of the 423 HARPS exposures. In each exposure, the core of each line --
the central 7 pixels, spanning $\approx$5.7\,\,km\,s$^{-1}$ -- was
fitted with a Gaussian model to determine the centroid wavelength. The
wavelength differences between line pairs in each exposure were then
computed, incorporating the corrections for the effects discussed
above \cite{Berke_2022a,Berke_2022b,materials}. We denote these pair
separations $\Delta v^i_{\rm raw}$ for pair $i$. In principle, they
can be compared to gauge any $\alpha$ variation between these 17 solar
twins. However, an analysis of 130 stars spanning a larger range in
$T_{\rm eff}$, [Fe/H] and $\log g$ (300\,K, 0.3\,dex and 0.4\,dex
around solar values, respectively), has shown that pair velocity
separation varies systematically with the stellar parameters,
typically by $\sim$60\,m\,s$^{-1}$ across this range
\cite{Berke_2022b,materials}.  We fit a quadratic model to those
correlations and use it to compute the expected line pair separation
for each star in our sample, denoting the resulting values
$\Delta v^i_{\rm model}$. We also incorporate an intrinsic
star-to-star scatter, $\sigma^i_{**} \approx 0$ to 15\,m\,s$^{-1}$
\cite{Berke_2022a,materials}. We then use $\Delta v^i_{\rm model}$ to
correct the observed separations for each individual star:
\begin{equation}\label{e:Deltavsep}
\Delta v^i_{\rm sep} = \Delta v^i_{\rm raw} - \Delta v^i_{\rm model}(T_{\rm eff}, {\rm [Fe/H]}, \log g)\,.
\end{equation}
For each line pair $i$, the value of $\sigma^i_{**}$ is the
systematic error in $\Delta v^i_{\rm sep}$; it is the typical absolute
value of the intrinsic deviation from the model.

The $\Delta v^i_{\rm sep}$ values have previously been calculated
\cite{Berke_2022a} from the HARPS solar twin exposures. For each line
pair in each solar twin, the velocity separation measurements from
multiple exposures were combined using a weighted mean, with outliers
excluded via an iterative process
\cite{Berke_2022b,materials}. Multiple exposures were available for 14
of the solar twins, allowing us to check for systematic errors as a
function of time. The optical fibres that feed light from the
telescope into HARPS were changed in mid-2015, resulting in large
calibration changes. Analysis of the pre- and post-fibre change epochs
separately -- including the determination of $\Delta v^i_{\rm model}$
-- has shown no evidence for systematic differences in
$\Delta v^i_{\rm sep}$ between them \cite{Berke_2022a}. We therefore
combined their weighted mean $\Delta v^i_{\rm sep}$ values. Three of
the 17 line pairs appear twice in each exposure, because they are in
the overlapping wavelength ranges of neighbouring diffraction
orders. We treat these two instances separately because we found
differences of $\sim$20\,m\,s$^{-1}$ between their
$\Delta v^i_{\rm raw}$ values, which we ascribe to optical distortions
within HARPS. Nevertheless, their weighted mean $\Delta v^i_{\rm sep}$
values show no systematic differences for our 17 stars, or the larger
sample \cite{Berke_2022b}, so we combine the $\Delta v^i_{\rm sep}$
value for two instances of a pair using a weighted mean.

Figure 2 shows our derived values of $\Delta\alpha/\alpha$ for each
star. The $\Delta v^i_{\rm sep}$ value for a line pair $i$ is
converted to $\Delta\alpha/\alpha$ using equations 1 and 2, and the
$Q$ coefficient calculations \cite{Dzuba_2022PhRvA.105f2809D}. For
each star plotted (Fig.\ 2), the $\Delta\alpha/\alpha$ values from all
pairs are consistent with each other, so they are combined using a
weighted mean. The weights in that process, and the final
uncertainties plotted (Fig.\ 2), include the statistical
uncertainties, derived from the SNR of the HARPS spectra, and
systematic errors that incorporate the star-to-star scatters for all
line pairs ($\sigma^i_{**}$) and a smaller contribution from the
uncertainties in the $Q$ coefficients. Because a line can be shared by
multiple pairs, its statistical and systematic uncertainties cause
correlated errors across those pairs; we used a Monte Carlo method to
compute the combined $\Delta\alpha/\alpha$ value and its statistical
and systematic uncertainty for each star \cite{materials}.

We find no variations in $\alpha$ between nearby solar twins
($<$50\,parsec), with a typical (median) uncertainty in
$\Delta\alpha/\alpha$ of $\approx$50\,parts per billion (ppb; adding
statistical and systematic errors in quadrature). The precision
reaches $\approx$30\,ppb for some stars, which is $\gsim$30 times more
precise than individual quasar absorption systems
\cite{Kotus_2017MNRAS.464.3679K,Murphy_2022A&A...658A.123M}. The
systematic error term dominates in these cases (Fig.\ 2), mainly due
to the intrinsic star-to-star scatter, $\sigma^i_{**} \approx 0$ to
15\,m\,s$^{-1}$ per line pair $i$. That is, the solar twins method
allows $\gsim$100 times more accuracy than comparison between lines in
individual white dwarfs or giant stars with their laboratory
counterparts
\cite{Hees_2020PhRvL.124h1101H,Hu_2021MNRAS.500.1466H}. The results
for the 17 stars are formally consistent with each other, with
$\chi^2 = 18.2$ around their weighted mean (16 degrees of freedom;
31\% probability of a larger value by chance alone), so there is no
evidence for additional systematic errors that are not accounted for
by $\sigma^i_{**}$.

Another study \cite{Berke_2022b} considered a variety of astrophysical
and instrumental effects that could cause spurious variation of
$\alpha$ between stars and/or account for $\sigma^i_{**}$. Apart from
those already corrected in our analysis (e.g.\ wavelength calibration
distortions), that study ruled out systematic error contributions from
line blending, pair separation, differences in line depth in a pair, 
transiting exoplanets or magnetic activity cycles of the target stars,
contamination of spectra by scattered moonlight, cosmic ray events or
charge transfer inefficiencies in the detector. However, they
estimated that variations in stellar rotation velocities or elemental
and isotopic abundances between stars may plausibly explain the size
of, and variations in, $\sigma^i_{**}$
\cite{Berke_2022b}. Nevertheless, they did not find specific evidence
for these effects with simple tests, even amongst the larger data
sample of stars used \cite{Berke_2022b}.

Combining the results from all 17 stars provides a weighted mean with
12\,ppb ensemble precision:
\begin{equation}
\left<\Delta\alpha/\alpha\right>_{\rm w} = 7 \pm 5_{\rm stat} \pm 11_{\rm sys}\,{\rm ppb}\,,
\end{equation}
where $\left<\Delta\alpha/\alpha\right>_{\rm w}$ and its 1$\sigma$
statistical uncertainty and systematic error were calculated from the
Monte Carlo simulations to account for the correlations between
results for different stars because they share common line pairs
\cite{materials}. The weights are the inverse variances from
quadrature addition of the statistical and systematic uncertainties in
$\Delta\alpha/\alpha$ for each star. The combined result (equation 3)
acts as an entirely empirical reference for stellar measurements of
$\alpha$. This, and the ability of our automatic analysis procedure to
recover shifts in $\alpha$ between stars, was tested by altering the
wavelength measurements for half our twins by amounts corresponding to
an $\alpha$ variation of 100\,ppb \cite{materials}. Re-running the
full analysis but removing these stars from the determination of
$\Delta v^i_{\rm model}$, we recovered an $86 \pm 19$\,ppb difference
between the shifted and unshifted twins. The discrepancy arises
because some measurements of shifted lines are excluded as outliers --
the shifts introduced are much larger than the total uncertainties
(including $\sigma^i_{**}$). This confirms that our analysis process
would still have detected any large ($\sim$100\,ppb) discrepancies
between some twins if they were present in the data.




\section*{Acknowledgments}
We thank Dainis Dravins for discussions about potential astrophysical systematic errors.

\noindent \textbf{Funding:} M.T.M., F.L.\ and C.L.\ acknowledge the support of the Australian Research Council through \textsl{Future Fellowship} grant FT180100194. V.A.D.\ and V.V.F.\ acknowledge the Australian Research Council for support through grants DP190100974 and DP200100150. OzGrav is funded by the Australian Government through the Australian Research Council Centres of Excellence funding scheme.

\noindent \textbf{Author contributions:} M.T.M.\ conceived the stellar twins technique, acquired funding, supervised the project, calculated the $\Delta\alpha/\alpha$ values and wrote the draft manuscript. D.A.B.\ wrote the software, performed the velocity shift measurements and analysis, and curated all data. F.L., C.F.\ and C.L.\ assisted with the methodology, analysis and validation of results. V.A.D.\ and V.V.F.\ calculated the sensitivity coefficients ($Q$). All authors reviewed and revised the manuscript.

\noindent \textbf{Competing interests:} The authors declare no competing interests.

\noindent \textbf{Data and materials availability:} This work is based on observations obtained from the ESO Science Archive Facility and collected at the European Southern Observatory under ESO programme(s) listed in Table S1. The observations are available from the ESO Science Archive facility: http://archive.eso.org/eso/eso\_archive\_main.html. Our software for measuring the line wavelengths, and computing the line pair separations and models is available on Zenodo \cite{VarConLib22}. Tables of the lines used in this study, their laboratory wavelengths, our measured and model offsets from those values, and our measured line pair separations, models and $\sigma^i_{**}$ values are available on Zenodo \cite{BerkeGithub22}. Our software for computing $\Delta\alpha/\alpha$ in this work is available on Zenodo \cite{MurphyGithub22}. The stellar parameters and our measured $\Delta\alpha/\alpha$ values and uncertainties are listed in Table S2.


\section*{Supplementary Materials}
Materials and Methods\\
Figs.\ S1 to S4\\
Tables S1 to S3\\
References \textit{(23--34)}


\clearpage


\begin{figure}
\begin{center}
\centerline{\includegraphics[width=0.95\columnwidth]{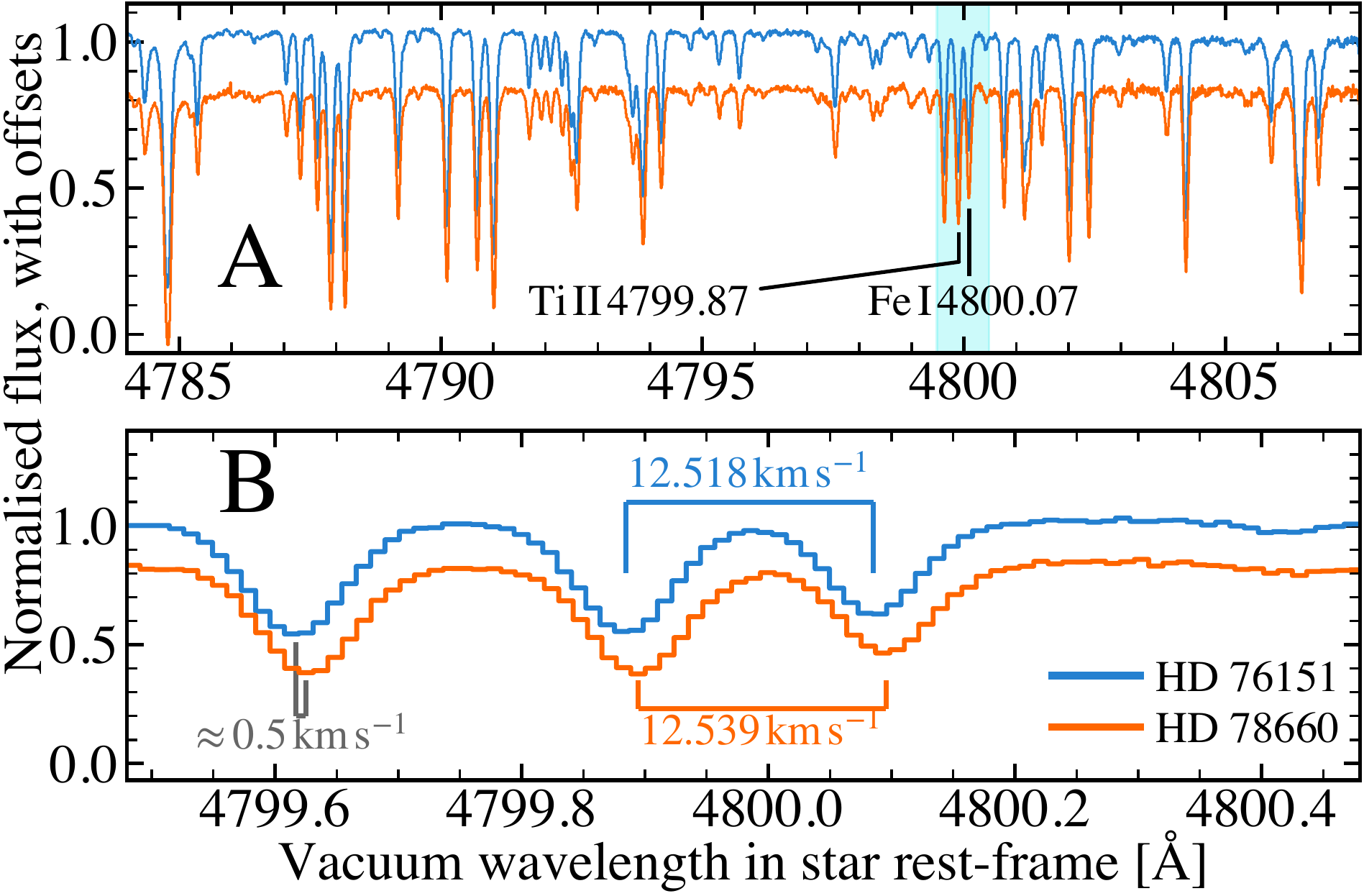}}
\caption{
\textbf{Example solar twin spectra.} (A) Small sections of continuum-normalised HARPS spectra of two solar twins from our sample, HD~76151 (blue) and HD~78660 (orange), with the latter shifted down by 0.2 for clarity. Black labels indicate an example line pair used to constrain $\alpha$ variation. This is the least-separated pair of the 17 in our analysis; the maximum separation of 800\,km\,s$^{-1}$ would span approximately half the width of this panel. The cyan shading indicates the region shown in panel B. (B) The measured velocity separations $\Delta v^i_{\rm raw}$, labelled in the figure, differ by 21\,m\,s$^{-1}$, before correction for their stellar parameters (equation 2). In both panels, the spectra are shown in the stars' rest frames; errors in their radial velocities are evident as a $\approx$0.5\,km\,s$^{-1}$ shift between them (labelled in grey in panel B). Our differential approach is insensitive to that offset.
}
\end{center}
\end{figure}

\begin{figure}
\begin{center}
\centerline{\includegraphics[width=0.90\columnwidth]{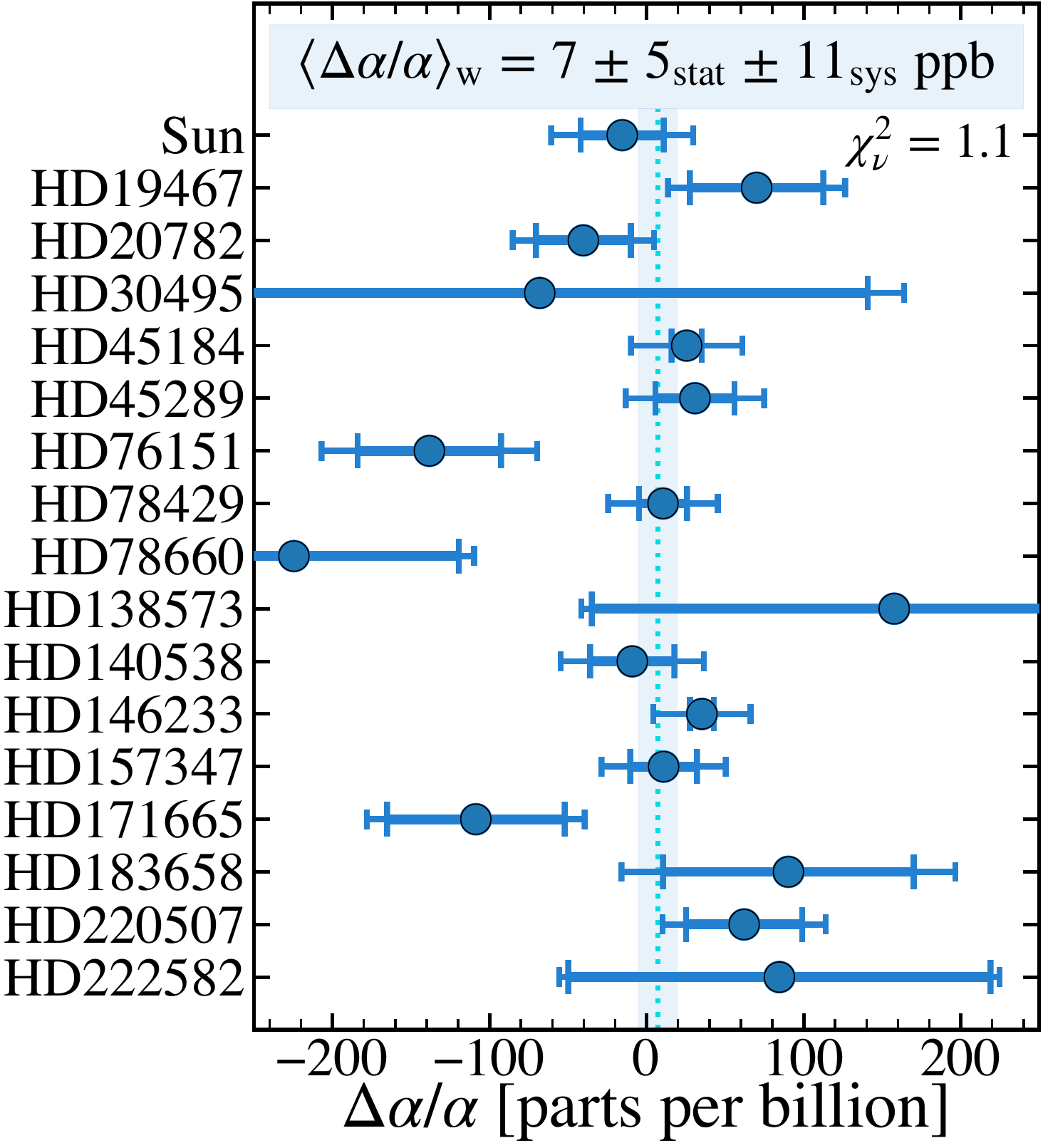}}
\caption{
\textbf{Fine-structure constant measurements.} The relative deviation of the fine-structure constant, $\Delta\alpha/\alpha$, is shown for each star in our sample. These values are averages of all 17 line pairs for each star. The inner large error bars indicate the 1$\sigma$ statistical uncertainties, dominated by the number of observations available of each star, while the outer small error bars combine the statistical and systematic uncertainties in quadrature. The weighted mean of the sample, $\left<\Delta\alpha/\alpha\right>_{\rm w}$, is shown as the dotted cyan line; the blue shaded region shows the combined 1$\sigma$ statistical and systematic uncertainties. $\chi^2_\nu=1.1$ is the reduced $\chi^2$ (per degree of freedom) of the individual measurements around the weighted mean.
}
\end{center}
\end{figure}

\clearpage

\begin{center}
{\Large Supplementary Materials for}

{\LARGE Strong limit on variations in the fine-structure constant from spectra of nearby Sun-like stars}

\bigskip 
\bigskip 

\baselineskip18pt

{\large Michael T. Murphy,$^{\ast}$ Daniel A. Berke, Fan Liu (刘凡), Chris Flynn,\\
Christian Lehmann, Vladimir A. Dzuba, Victor V. Flambaum}
\end{center}

\baselineskip14pt

\bigskip 
\bigskip 

\noindent\textbf{This PDF file includes:}\\
\indent Materials and Methods\newline
\indent Figs.\ S1 to S4\newline
\indent Tables S1 to S3\newline

\clearpage

\section*{Materials and Methods}

The full details of the solar twins analysis are presented elsewhere
\cite{Berke_2022a}, as is an application to a larger sample of solar
analogues (broader than solar twins) \cite{Berke_2022b}. We adopt
their modelling of raw line pair separations as functions of the main
stellar atmospheric parameters: effective temperature $T_{\rm eff}$,
iron metallicity [Fe/H], and surface gravity $\log g$
\cite{BerkeGithub22}. Here we summarise that analysis, and provide
further details about the conversion of the pair velocity separations
to $\Delta\alpha/\alpha$.

\subsection*{Spectral line selection}

An initial catalogue of 8843 lines in the Sun's spectrum was drawn
from the Belgian Repository of fundamental Atomic data and Stellar
Spectra (BRASS) \cite{Laverick_2018A&A...612A..60L}. Most lines (90\%)
were excluded because they fell within 100\,km\,s$^{-1}$ of telluric
lines or within 3.5 HARPS resolution elements (9.1\,km\,s$^{-1}$) of
another stellar line. The former tolerance represents the maximum
velocity shift possible for lines in our sample of stars, which have
radial velocities of $\pm$70\,km\,s$^{-1}$, given the Earth's annual
barycentric velocity variation of $\pm$30\,km\,s$^{-1}$. All telluric
lines that absorb $\ge$0.1\% of the continuum were avoided to ensure
that the fitted centroid of any individual stellar line of interest
was affected by, at most, $\approx$30\,m\,s$^{-1}$ in a single
observation. Of the remaining 849 lines, 7\% were removed because they
were very weak or near saturation ($<$15\% or $>$90\% absorption
depth, respectively) as measured in a HARPS spectrum of Vesta
\cite{Haywood_2016MNRAS.457.3637H}. Most of the remaining 783 lines
(79\%) were then excluded because they appeared to be blended with
other stellar lines at HARPS's resolving power ($R \approx 115,000$)
\cite{Mayor_2003Msngr.114...20M}. Only close pairs of lines (within
800\,km\,s$^{-1}$), which had similar absorption depths (within 20\%),
were then selected. This left 164 lines, contributing to 229 pairs for
study \cite{Berke_2022b}. We adopted calculated sensitivity
coefficients, $q$\cite{Dzuba_2022PhRvA.105f2809D}, for 22 of these
lines, where $Q\equiv q/\omega_0$, with $\omega_0$ the rest-frame
wavenumber (see equation 1). These 22 lines, which we combine to form
17 pairs (some pairs share common lines), constitute the spectral line
sample for which pair velocity separations were measured
\cite{Berke_2022a}. We calculate their $\Delta\alpha/\alpha$ values
below.

\subsection*{Stellar sample and spectroscopic data selection}

Beyond the requirement that stars in our sample must have been
observed with HARPS, our main goal in selecting a stellar sample was
to reduce bias in our study of solar twins. Our method requires a
model of the variation of line pair separations with stellar
parameters. It is therefore desirable for the sample to uniformly
sample the three-dimensional stellar parameter space to avoid
biases. For example, if the sample was concentrated on the solar twins
range, with sparser sampling outside it, the models of variations in
pair separation would have been unduly weighted towards the
twins. Therefore we selected the sample from an approximately
homogenous starting catalogue \cite{Casagrande_2011A&A...530A.138C}.

We selected stars from the 14139 F and G dwarfs in the
Geneva--Copenhagen Survey \cite{Nordstrom_2004A&A...418..989N}. Using
improved photometric stellar parameter estimates
\cite{Casagrande_2011A&A...530A.138C}, this catalogue contains 711
`solar type' stars according to a standard definition
\cite{Berke_2022b}, 200 of which had at least one exposure with
signal-to-noise ratio $200 \le {\rm SNR} \le 400$ per
0.8\,km\,s$^{-1}$ HARPS pixel in the ESO archive \cite{ESOarchive}. To this
sample we added the Sun, observed indirectly via reflection from the
asteroid Vesta; to increase the number of Vesta exposures used in our
analysis, we only required a threshold SNR of 150 per pixel. Of these
201 stars, 136 had exposures for which corrections to the wavelength
calibration for non-uniform detector pixel sizes were available
\cite{Coffinet_2019A&A...629A..27C}. Finally, to reduce the number of
lines falling near telluric features, we removed 6 stars with absolute
radial velocities $>$70\,km\,s$^{-1}$ (see above).

These 130 `solar type' stars were used to model the variation of each
line pair's velocity separation with stellar atmospheric parameters
\cite{Berke_2022b,BerkeGithub22}. These parameters differ from the
solar values \cite{Prsa_2016AJ....152...41P} (i.e.\
$T_{\rm eff}=5772$\,K, $\textrm{[Fe/H]}=0.0$\,dex, $\log g=4.44$\,dex,
for $g$ in units of cm\,s$^{-1}$) over the following ranges:
$-350 \le \Delta T_{\rm eff} \le 450$\,K,
$-0.65 \le \textrm{[Fe/H]} \le 0.40$\,dex and
$-0.3 \le \log g \le 0.2$. The pair velocity separations have been
measured for the 18 `solar twins' in this sample \cite{Berke_2022a},
i.e.\ where the stellar parameters differed by 100\,K, 0.1\,dex and
0.2\,dex from the solar values, respectively. We exclude the results
from one of these stars (HD~1835) because the single HARPS exposure
available provided much weaker constraints on $\Delta\alpha/\alpha$
than the other stars ($\Delta\alpha/\alpha\approx2300$\,ppb with total
uncertainty $\gsim$1200\,ppb). Table S1 provides the ESO Programme
identification codes for the HARPS exposures of the 17 stars studied
here.

\begin{table}
\renewcommand\thetable{S1}
\caption{\textbf{Origin of the HARPS exposures}. The first column provides the ESO Programme identification codes, with the principal investigator named in the second column. The third column provides the names of the objects observed.}
\label{t:eso}
\begin{center}
\begin{tabular}{lcl}\hline
Project ID & PI Name         & Objects \\\hline
60.A-9036  & ESO$^{\rm a}$    & HD~45184, HD~183658, Vesta \\
072.C-0488 & M.\ Mayor       & HD~19467, HD~20782, HD~45184, HD~45289, HD~76151, \\
           &                 & HD~78429,  HD~146233, HD~157347, HD~171665, HD~183658, \\
           &                 & HD~220507, HD~222582 \\
075.C-0332 & C.\ H. F.\ Melo & HD~78660, HD~140538 \\
077.C-0364 & M.\ Mayor       & HD~146233, HD~171665, HD~183658 \\
088.C-0323 & A.\ Cameron     & Vesta \\
091.C-0936 & S.\ Udry        & HD~45184 \\
183.C-0972 & S.\ Udry        & HD~19467, HD~20782, HD~45184, HD~45289, HD~76151, \\
           &                 & HD~146233, HD~157347, HD~171665, HD~183658, HD~220507, \\
           &                 & HD~222582 \\
183.D-0729 & M.\ Bazot       & HD~78660, HD~146233, HD~220507 \\
188.C-0265 & J.\ Mel\'endez  & HD~19467, HD~30495, HD~45289, HD~78660, HD~138573, \\
           &                 & HD~140538, HD~146233, HD~157347, HD~183658, HD~220507 \\
192.C-0852 & S.\ Udry        & HD~19467, HD~20782, HD~45184, HD~78429, HD~146233, \\
           &                 & HD~157347, HD~183658, HD~220507, HD~222582 \\
196.C-1006 & S.\ Udry        & HD~20782, HD~78429, HD~146233, HD~220507 \\
198.C-0836 & R.\ Diaz        & HD~45184, HD~78429, HD~146233, HD~220507 \\\hline
\end{tabular}
\end{center}
$^{\rm a}$ESO uses the project ID 60.A-9036 to indicate observations undertaken for calibration or engineering purposes.
\end{table}

\subsection*{Data processing and line wavelength measurement}

The HARPS spectra were processed automatically by the HARPS Data
Reduction System (DRS) \cite{Rupprecht_2004SPIE.5492..148R}. We used
intermediate data products to produce a flux uncertainty array for
each echelle order, following the approach of a previous study
\cite{Dumusque_2018A&A...620A..47D}, to determine an uncertainty
estimate for the centroid wavelength of each spectral feature. The
flux and uncertainty arrays are not merged to form a one dimensional
spectrum. Instead, the observed wavelength of each line is measured
from the two dimensional spectrum of the one or two echelle orders in
which it falls.

Line positions were determined by fitting a Gaussian model to the
central 7 pixels, equivalent to the central
$\approx$5.7\,\,km\,s$^{-1}$ of the line, which reduces line asymmetry
effects
\cite{Dravins_1982ARA&A..20...61D,Dravins_2008A&A...492..199D}. For
each pixel, the Gaussian profile is averaged, rather than evaluated
only at its centre, to account for the non-uniform flux distribution
across its spectral dimension \cite{Berke_2022b}. The resulting
centroid wavelength is already corrected (by the DRS) for the
barycentric motion of the Earth during the observation and the
non-uniform detector pixel sizes. An additional wavelength correction
is applied to account for inaccuracies in the standard thorium--argon
lamp wavelength calibration \cite{Berke_2022b}: a low-order polynomial
model fitted to a small number of calibration lines on each echelle
order leaves $\sim$5\,m\,s$^{-1}$ distortions in the calibration, even
after correction for the non-uniform pixel sizes
\cite{Milakovic_2020MNRAS.493.3997M}.

Before the observed line wavelengths from a single exposure are used
to measure pair separations, their offsets from their respective
laboratory wavelengths are used for identifying if any of them are
outliers. This allows the rejection of lines in which a cosmic ray
event affected the flux recorded in one or more of the seven fitted
pixels. The offset for a single line is strongly correlated with the
primary three stellar parameters, $T_{\rm eff}$, [Fe/H] and $\log g$,
so it was modelled as a quadratic function in each of these parameters
\cite{Berke_2022a,Berke_2022b}. As an example, Fig.\ S1 shows the
variation in the offset for one line, Fe\,{\sc i}\,6138.313, as a
function of metallicity. The line offset measurements and models for
all stars are provided elsewhere \cite{Berke_2022b,BerkeGithub22}. For
each exposure of each star, individual lines whose offsets deviated
from the model by more than 3 times their statistical uncertainties
were rejected. This was determined in an iterative procedure in which
the best-fitting radial velocity for the exposure was simultaneously
measured. This procedure employed all 164 lines that satisfied the
line selection criteria discussed above.

\renewcommand\thefigure{S1}
\begin{figure}
\begin{center}
\centerline{\includegraphics[width=0.67\columnwidth]{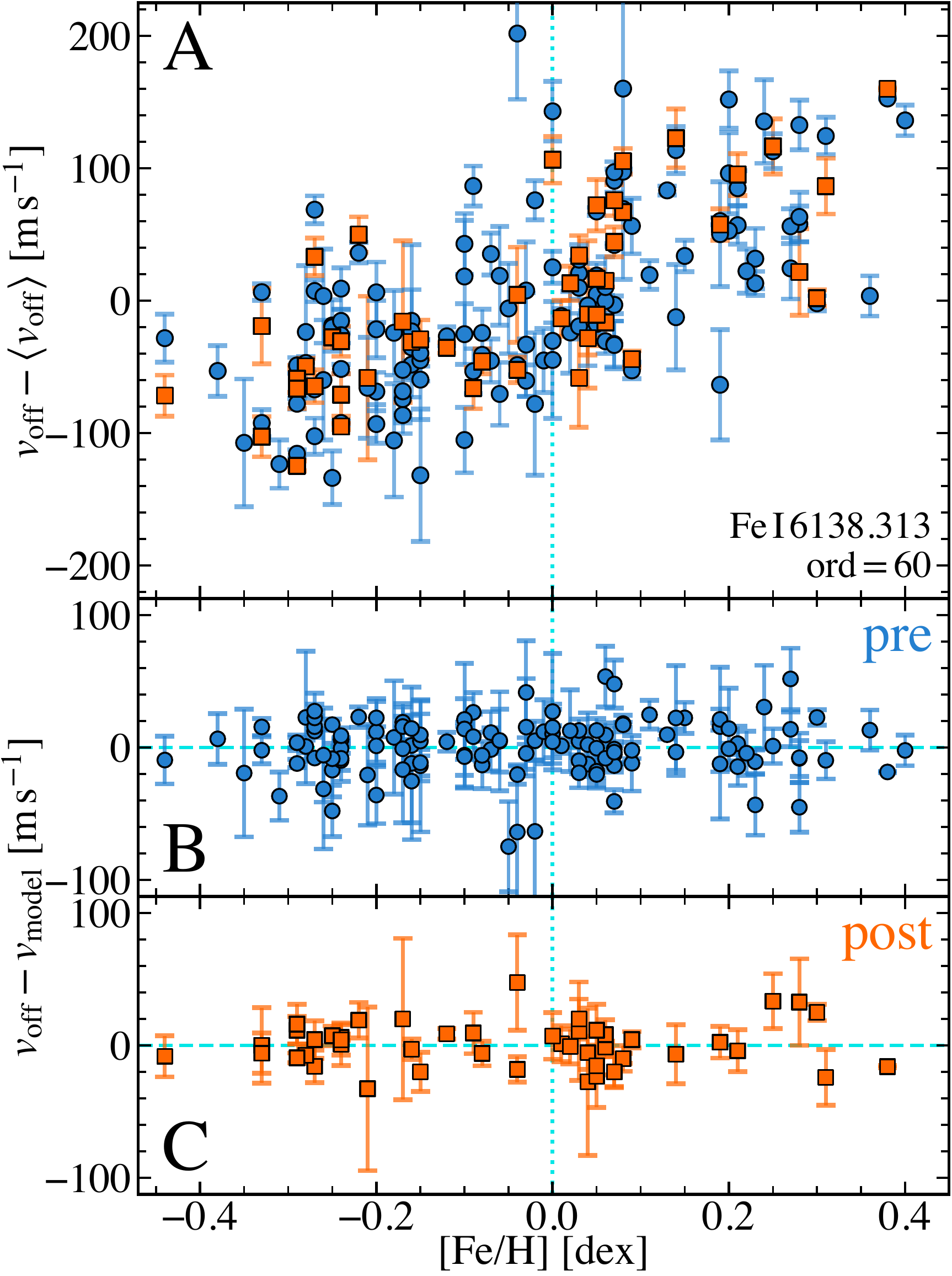}}
\caption{
\textbf{Example of variation in line offset with metallicity, [Fe/H].} (A) Deviation of the measured line offset from its laboratory wavelength, $\Delta v_{\rm off}$, from the mean value, $\left<\Delta v_{\rm off}\right>$, in 130 `solar type' stars \cite{Berke_2022b} for the line Fe\,{\sc i}\,6138.313 in echelle order 60. Blue circles and orange squares indicate measurements from spectra observed in the pre- and post-fibre change epochs, respectively. Error bars indicate the 1$\sigma$ statistical uncertainties derived from the signal-to-noise ratio of the spectra. (B) Differences between the $v_{\rm off}$ values and the quadratic model, $v_{\rm model}$, calculated for the pre-fibre change measurements in panel A \cite{Berke_2022b,BerkeGithub22}. The dashed horizontal cyan line indicates zero deviation from the model. (C) Same as for panel B but for the post-fibre change measurements. The dotted vertical cyan line in all panels indicates the solar metallicity. Stars for which all measurements of this line were rejected as outliers appear in panel A but not in the lower two panels.
}
\end{center}
\end{figure}

\subsection*{Pair separation models}

Convective motions in stellar atmospheres shift and skew the shapes
of stellar lines
\cite{Dravins_1982ARA&A..20...61D,Dravins_2008A&A...492..199D}. More
photons are emitted from hotter, up-welling convective cells
(granules) than colder, sinking ones, which shifts stellar lines
bluewards from laboratory wavelengths. The blueshift varies with depth
in the photosphere, as do the local conditions (temperature and
pressure) which affect the line-width. Each observed stellar line is
the cumulative result of differing amounts of absorption over a range
of depths, so the line profile is asymmetric, with the degree of
asymmetry varying as a function of the line's absorption depth (i.e.\
the line bisector is not at a constant offset from the laboratory
wavelength). The shift and asymmetry differs for each line because the
amount of absorption, at a given depth in the photosphere, depends on
the atomic physics of that line, e.g.\ the ionisation level, the
line's initial energy state and quantum mechanical probability. Our
solar twins approach attempts to reduce these effects by comparing the
separations between the same pair of lines in very similar stars. The
astrophysical shifts and asymmetries will also depend on the physical
conditions in the photosphere, characterised by the stellar
atmospheric parameters $T_{\rm eff}$, [Fe/H] and $\log g$.

We therefore expect variations in the observed separation between
pairs of stellar lines as functions of these parameters. These are
likely to be of a similar magnitude, though somewhat smaller than, the
astrophysical shifts and asymmetries themselves, which are observed to
be up to $\sim$700\,m\,s$^{-1}$ in individual lines
\cite{Dravins_2008A&A...492..199D,GonzalezHernandez_2020A&A...643A.146G}. The
larger, `solar type' sample of 130 stars \cite{Berke_2022b} showed
that the raw pair separation -- $\Delta v^i_{\rm raw}$ in equation (2)
-- varies with the stellar parameters. As an example, one line pair
has a separation which varies by $\sim$400\,m\,s$^{-1}$ over the
metallicity range $-0.4 < \textrm{[Fe/H]} < 0.4$\,dex
[\cite{Berke_2022b}, their figure 15]. While this was noted as larger
than typical, it was not unrepresentative. For the 17 pairs studied in
this paper, the variation is $\sim$150\,m\,s$^{-1}$ at most. Fig.\ S2A
shows an example where the variation is $\sim$100\,m\,s$^{-1}$ over
the same metallicity range, and Fig.\ S3 shows the variation for each
pair as a function of all three stellar parameters ($T_{\rm eff}$,
[Fe/H] and $\log g$).

The variation in pair separation was found to be a low-order,
generally monotonic function of the stellar parameters for almost all
the 229 pairs included in the previous study \cite{Berke_2022b},
including all of the 17 pairs used in this paper. This implies that we
can use an empirical model for $\Delta v^i_{\rm model}$ in equation
(2), for each pair $i$, to provide a local reference for the pair
separation for any solar analogue star. We found that a quadratic
model adequately tracked the variation of $\Delta v^i_{\rm raw}$ with
$T_{\rm eff}$, [Fe/H] and $\log g$. Incorporating this model allows us
to compare stars which are not perfect twins of each other, having
different stellar parameters.

\renewcommand\thefigure{S2}
\begin{figure}
\begin{center}
\centerline{\includegraphics[width=0.67\columnwidth]{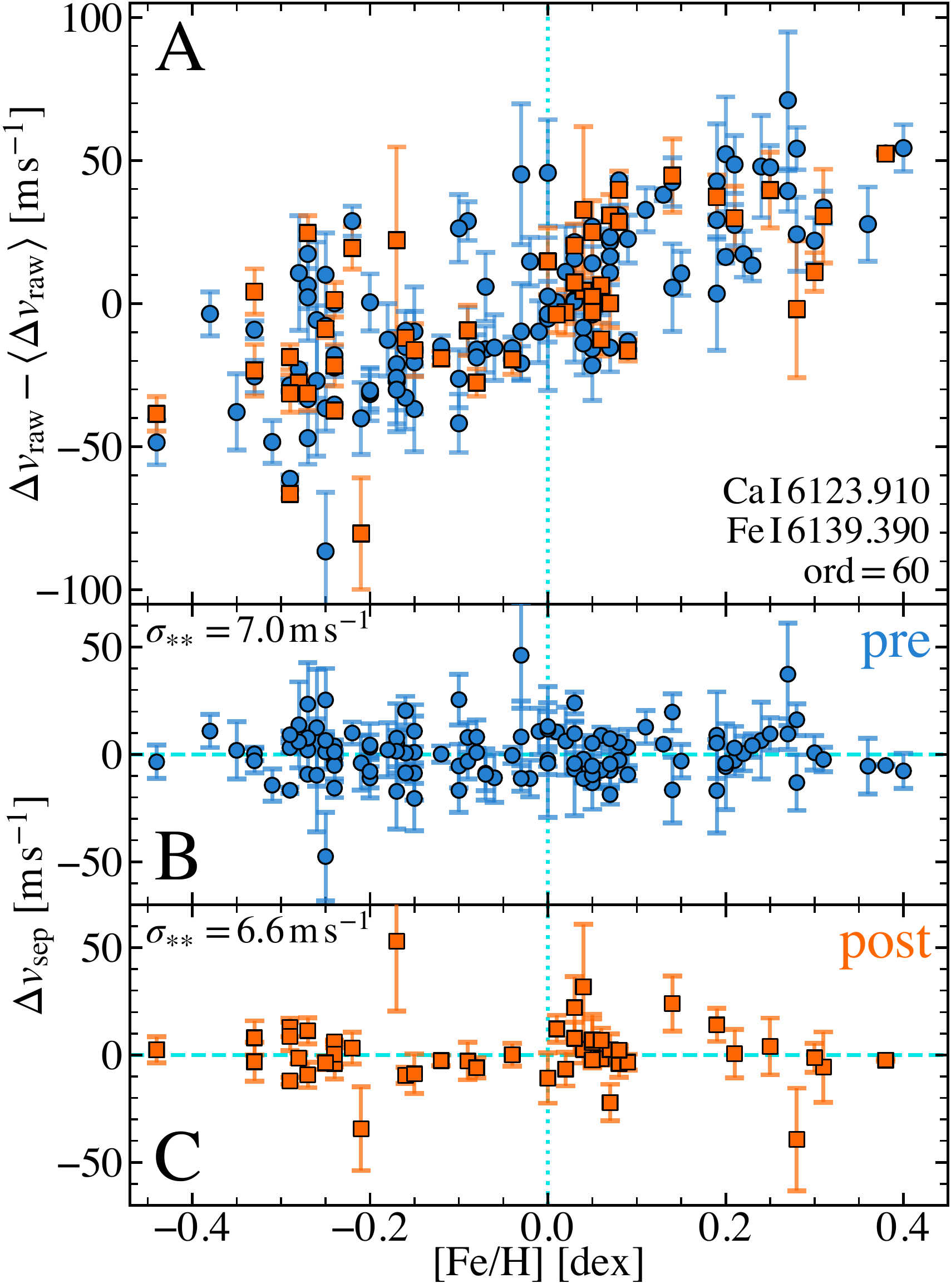}}
\caption{
\textbf{Example of variation in line pair separation with metallicity, [Fe/H].} (A) Deviation of the measured pair separations, $\Delta v_{\rm raw}$, from their mean value, $\left<\Delta v_{\rm raw}\right>$, in 130 `solar type' stars \cite{Berke_2022b} for the line pair and echelle order labelled in the figure: Ca\,{\sc i}\,6123.910 and Fe\,{\sc i}\,6139.390 in order 60. Blue circles and orange squares indicate measurements from spectra observed in the pre- and post-fibre change epochs, respectively. Error bars indicate the 1$\sigma$ statistical uncertainties derived from the signal-to-noise ratio of the spectra. (B) Values of $\Delta v_{\rm sep}$ calculated for the pre-fibre change measurements in panel A, i.e.\ the deviation of $\Delta v_{\rm raw}$ from the quadratic model of its variation with all three stellar parameters, $T_{\rm eff}$, [Fe/H] and $\log g$ (equation 2). The star-to-star scatter, $\sigma_{**}$ labelled in the figure, measures the additional scatter of $\Delta v_{\rm sep}$ in excess of that expected from the statistical uncertainties. The dashed horizontal cyan line indicates zero deviation from the model. (C) Same as for panel B but for the post-fibre change measurements. The dotted vertical cyan line in all panels indicates the solar metallicity.
}
\end{center}
\end{figure}

\renewcommand\thefigure{S3}
\begin{figure}
\begin{center}
\vbox{
\centerline{\includegraphics[width=0.90\columnwidth]{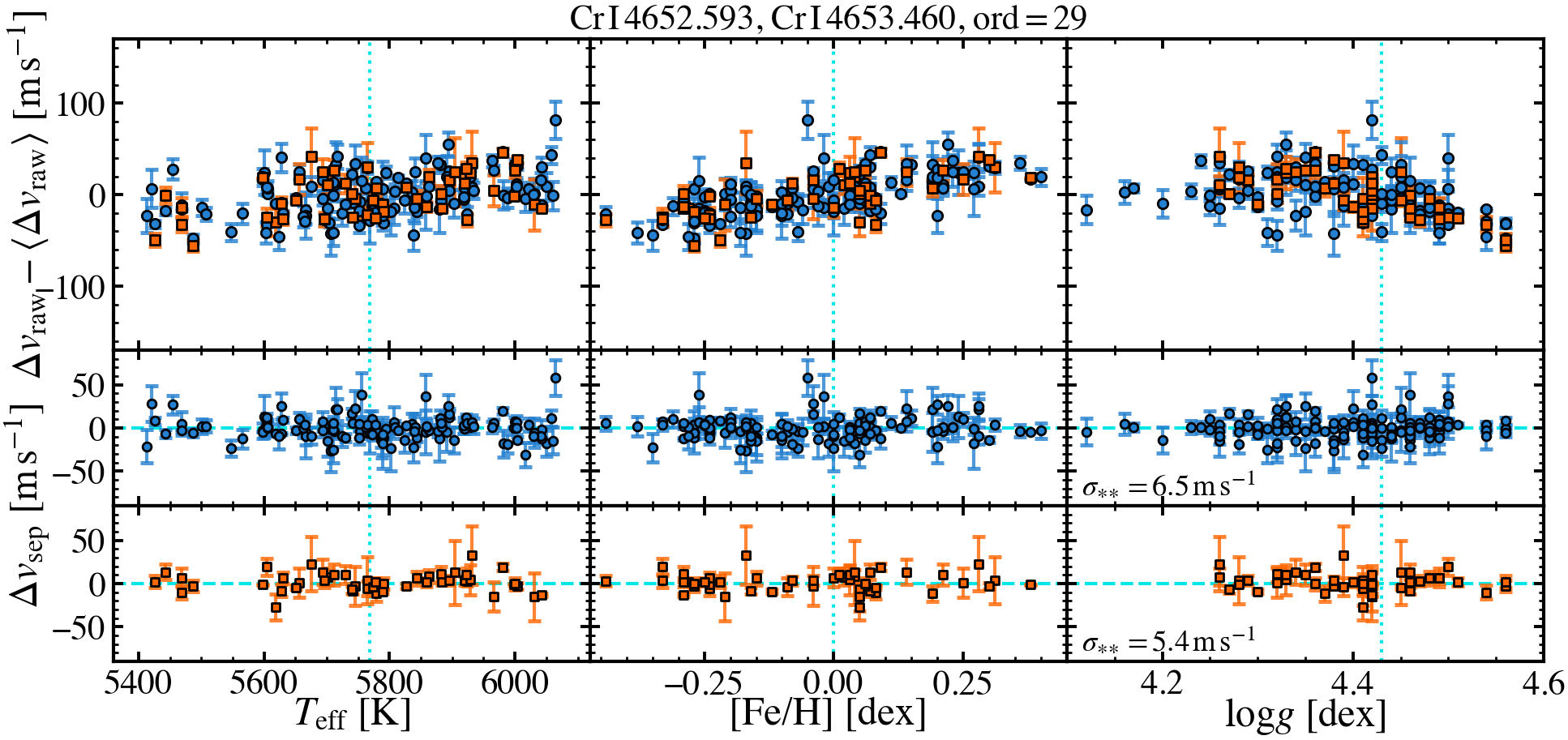}}
\centerline{\includegraphics[width=0.90\columnwidth]{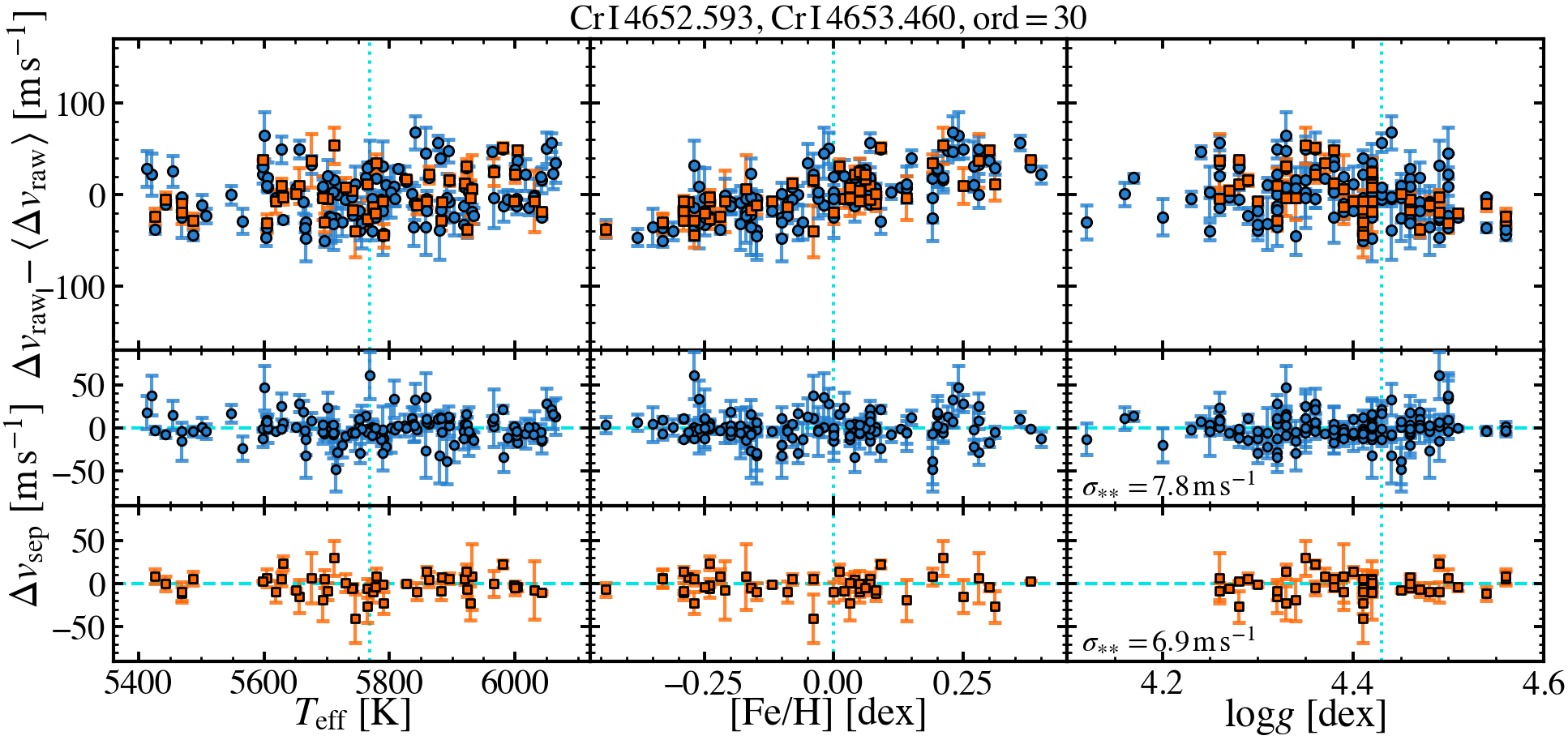}}
}
\caption{
\textbf{Variation in line pair separations.} Same as for Fig.\ S2 but for all three stellar parameters ($T_{\rm eff}$, [Fe/H] and $\log g$) and the pairs and echelle orders labelled above the figures.
}
\end{center}
\end{figure}

\renewcommand\thefigure{S3~(continued)}
\begin{figure}
\begin{center}
\vbox{
\centerline{\includegraphics[width=0.90\columnwidth]{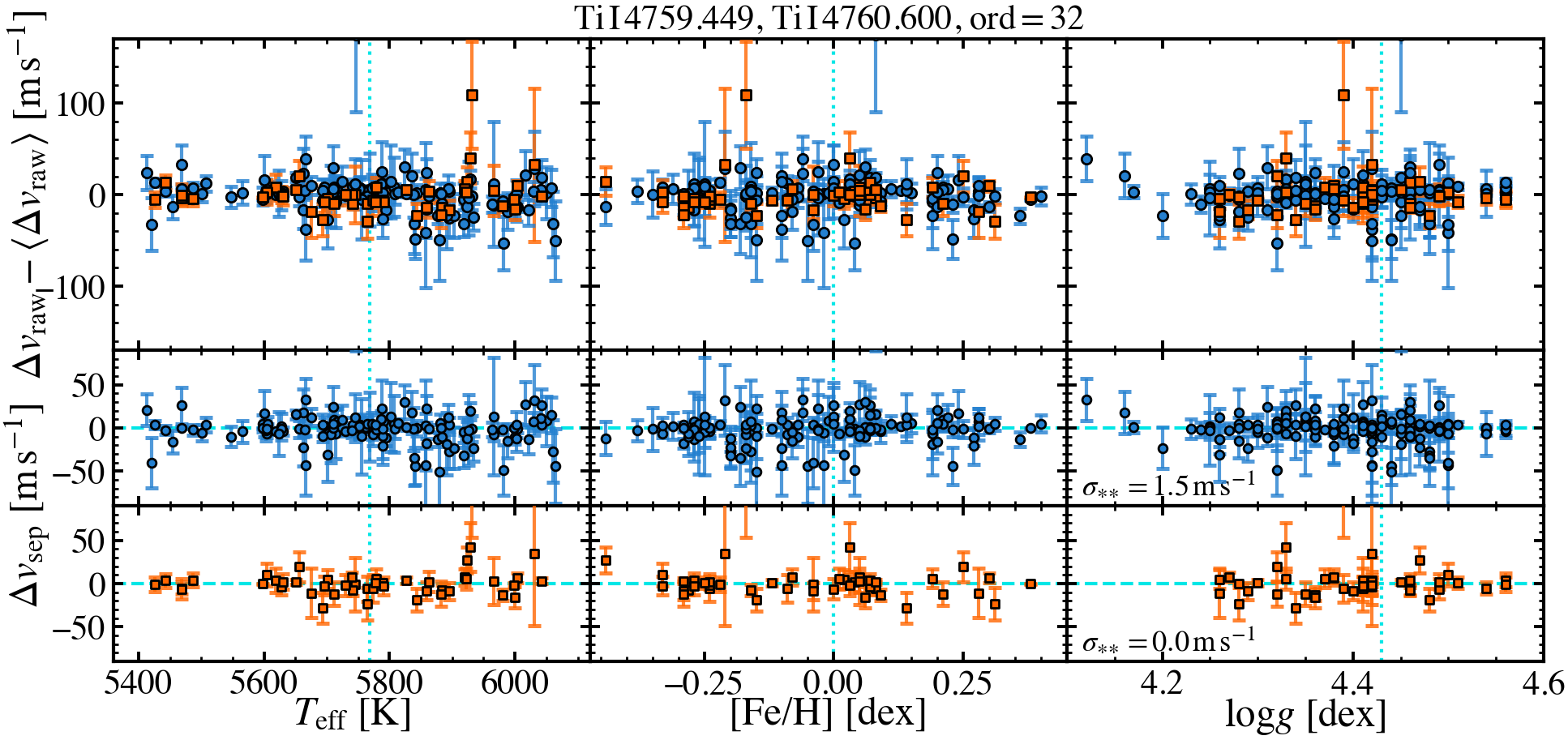}}
\centerline{\includegraphics[width=0.90\columnwidth]{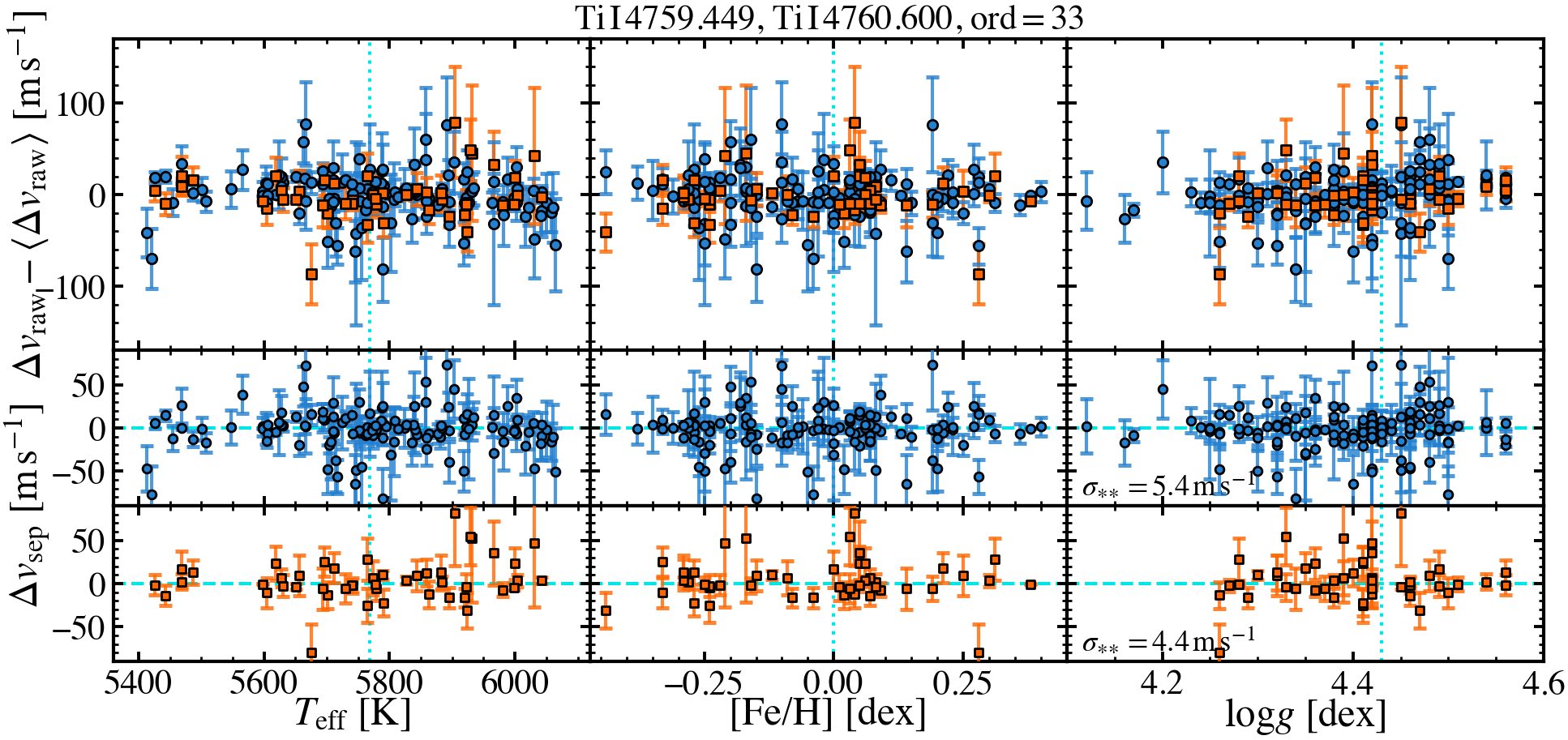}}
\centerline{\includegraphics[width=0.90\columnwidth]{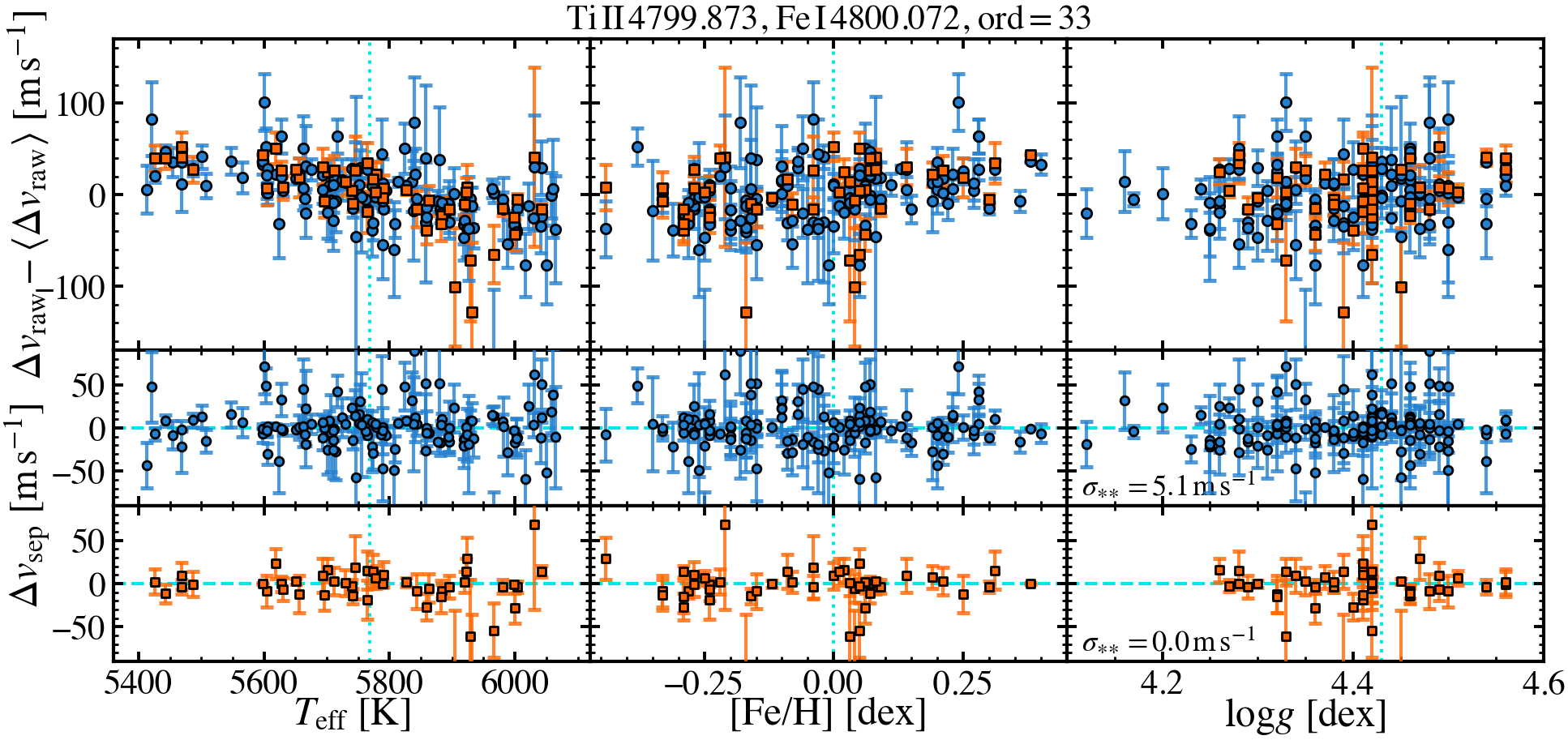}}
}
\caption{
\textbf{Variation in line pair separations.}
}
\end{center}
\end{figure}

\begin{figure}
\begin{center}
\vbox{
\centerline{\includegraphics[width=0.90\columnwidth]{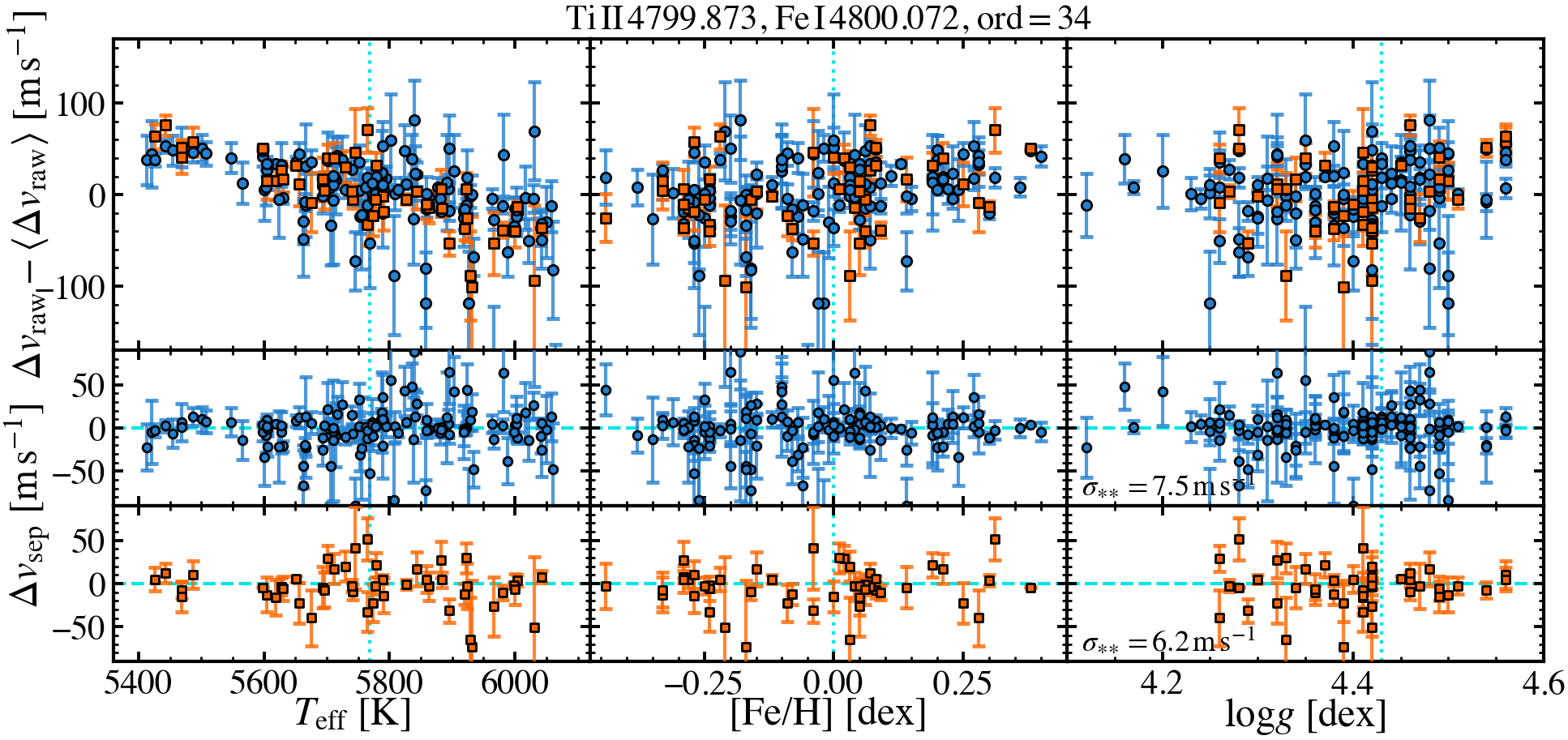}}
\centerline{\includegraphics[width=0.90\columnwidth]{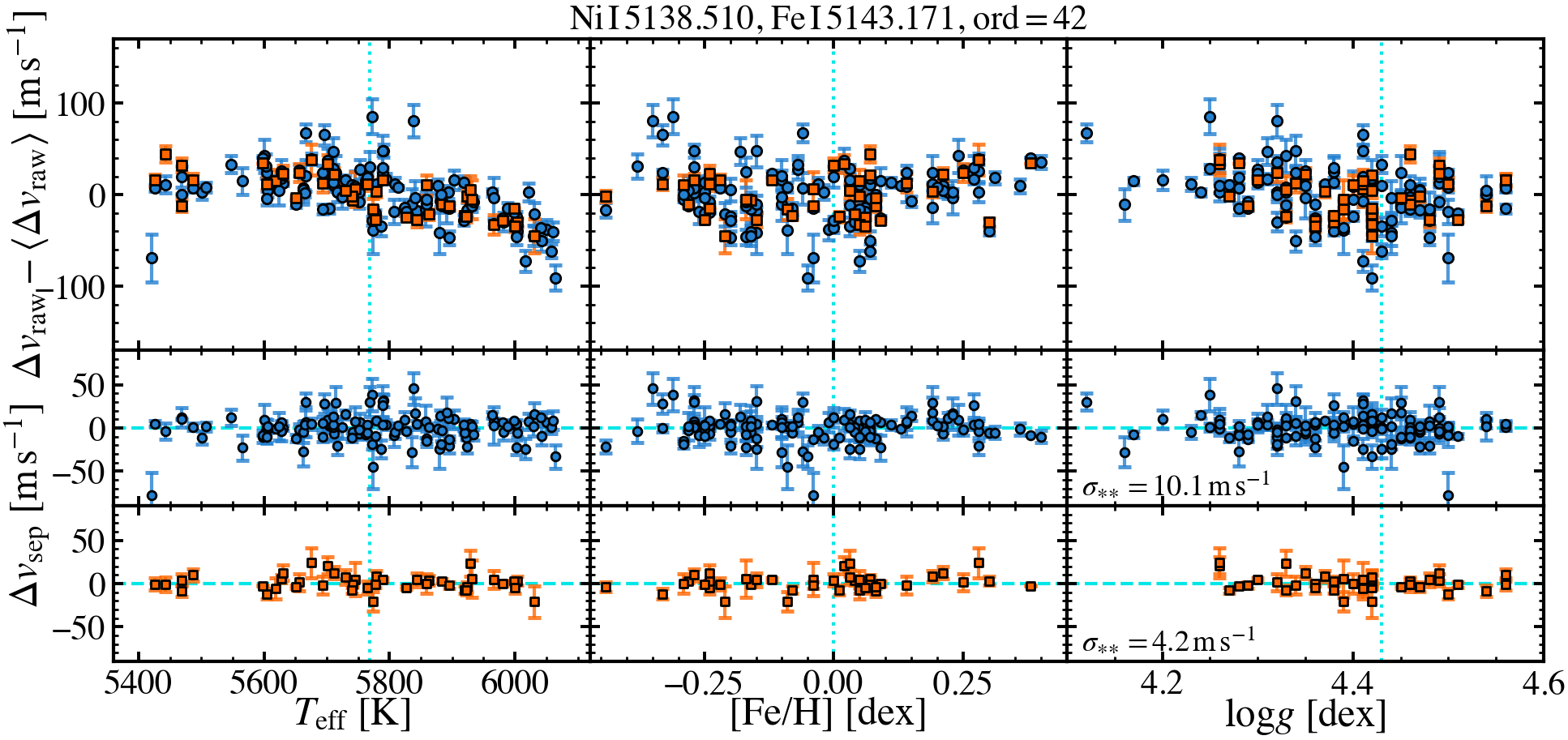}}
\centerline{\includegraphics[width=0.90\columnwidth]{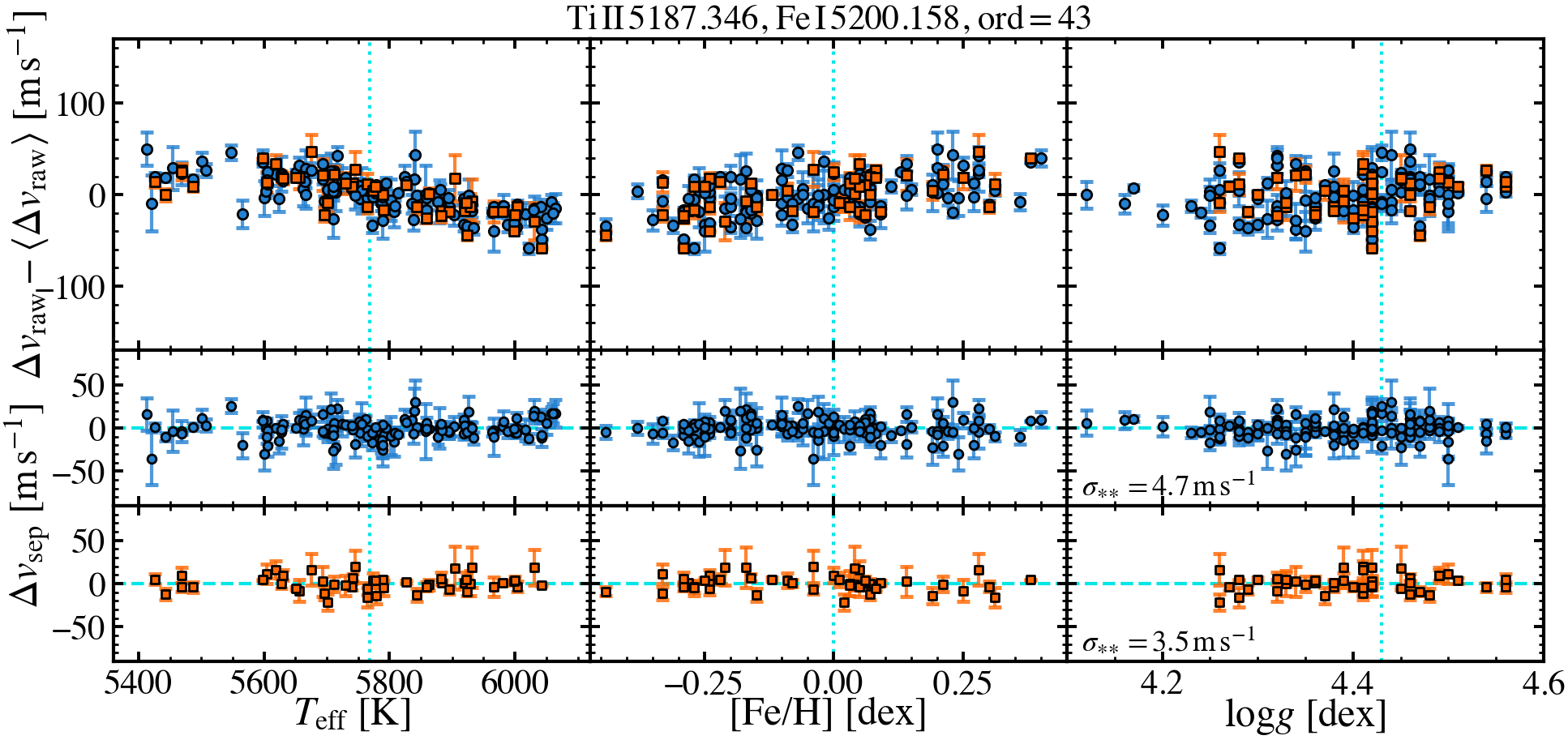}}
}
\caption{
\textbf{Variation in line pair separations.}
}
\end{center}
\end{figure}

\begin{figure}
\begin{center}
\vbox{
\centerline{\includegraphics[width=0.90\columnwidth]{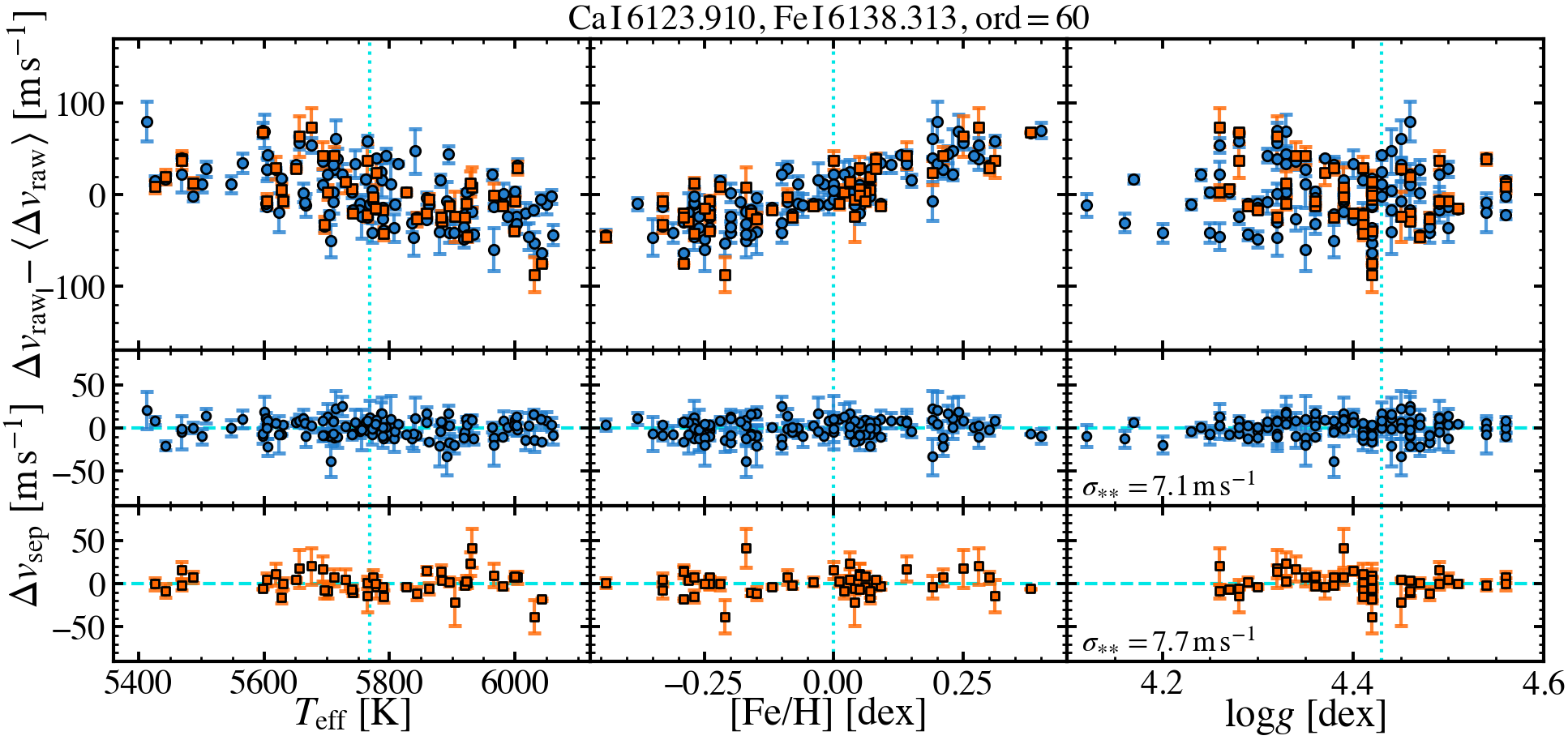}}
\centerline{\includegraphics[width=0.90\columnwidth]{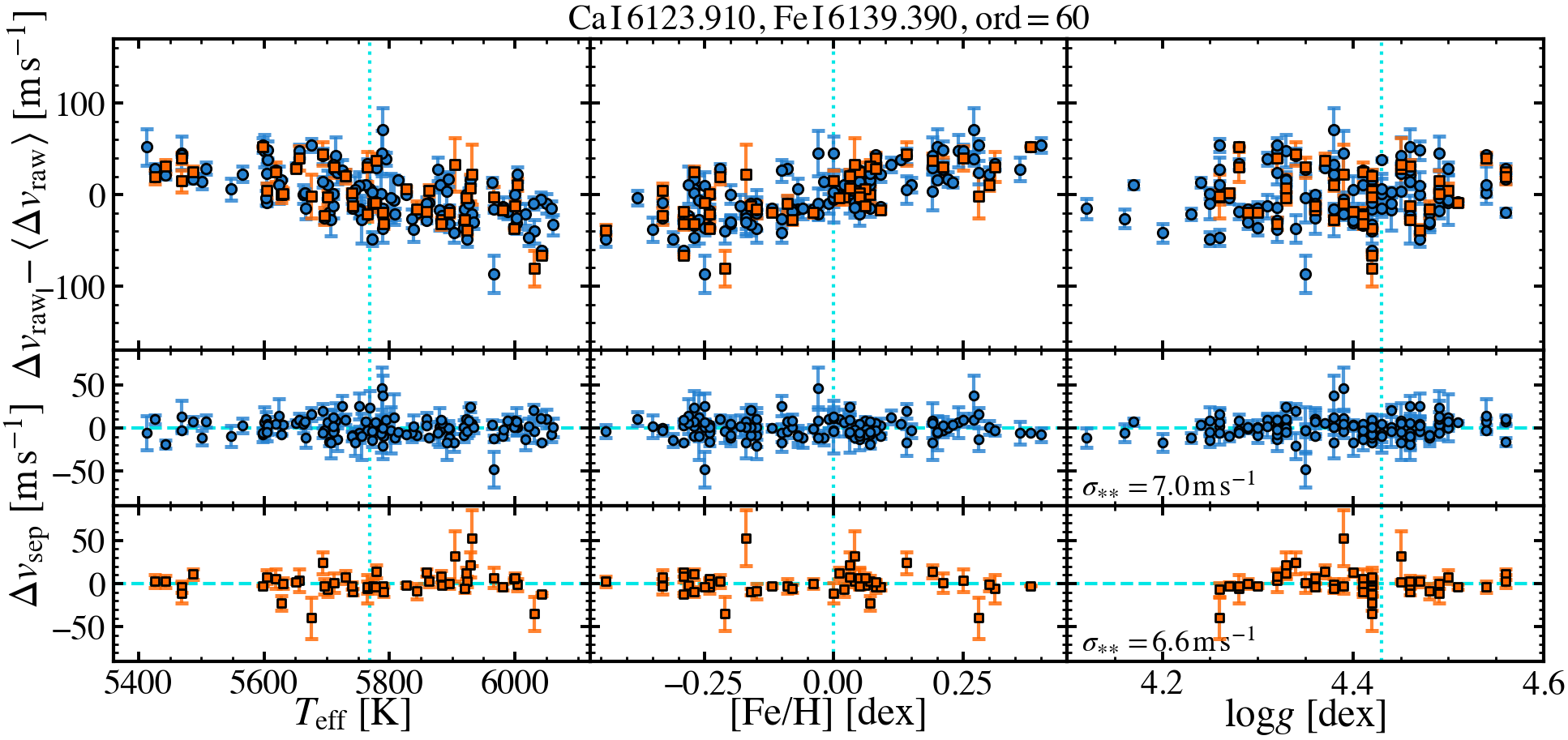}}
\centerline{\includegraphics[width=0.90\columnwidth]{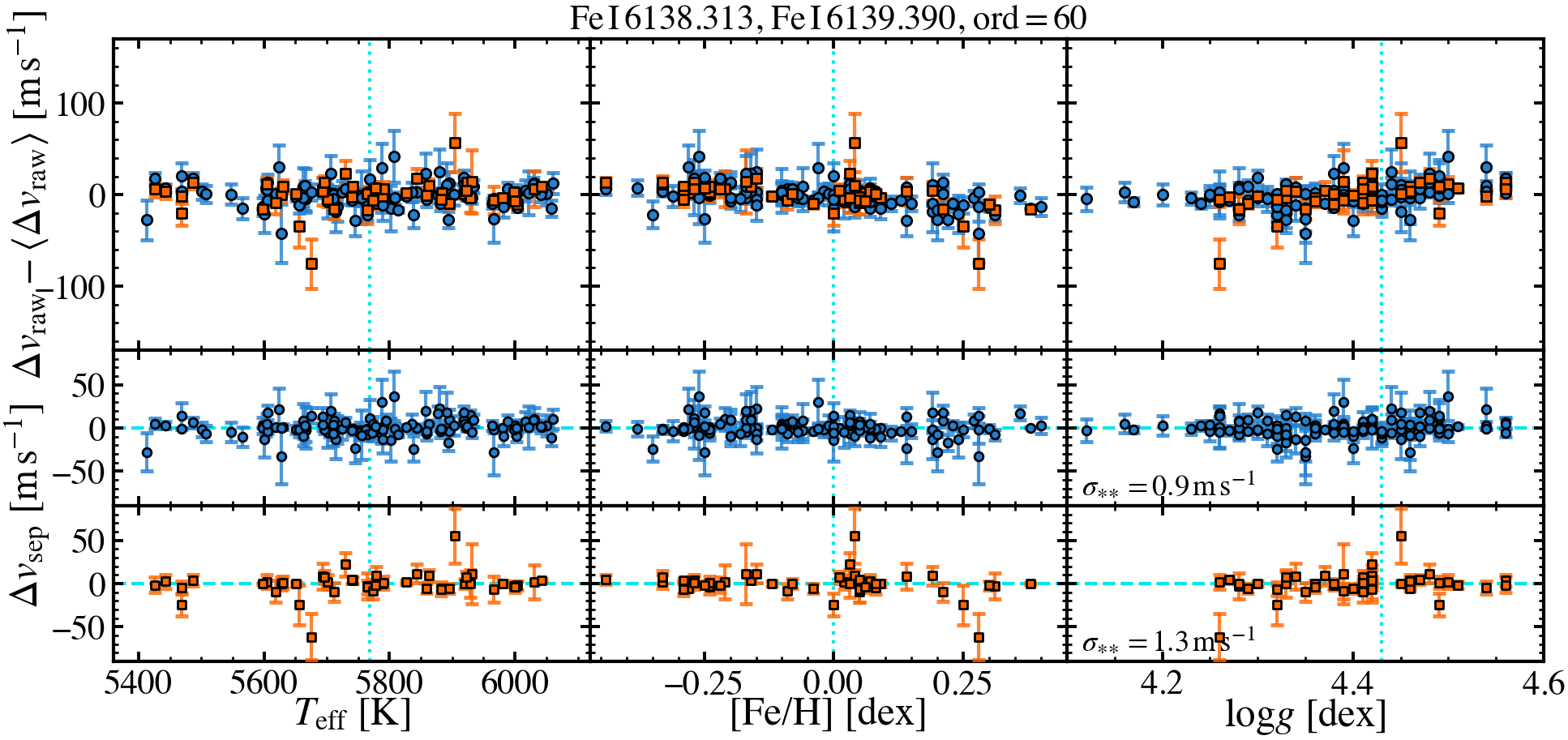}}
}
\caption{
\textbf{Variation in line pair separations.}
}
\end{center}
\end{figure}

\begin{figure}
\begin{center}
\vbox{
\centerline{\includegraphics[width=0.90\columnwidth]{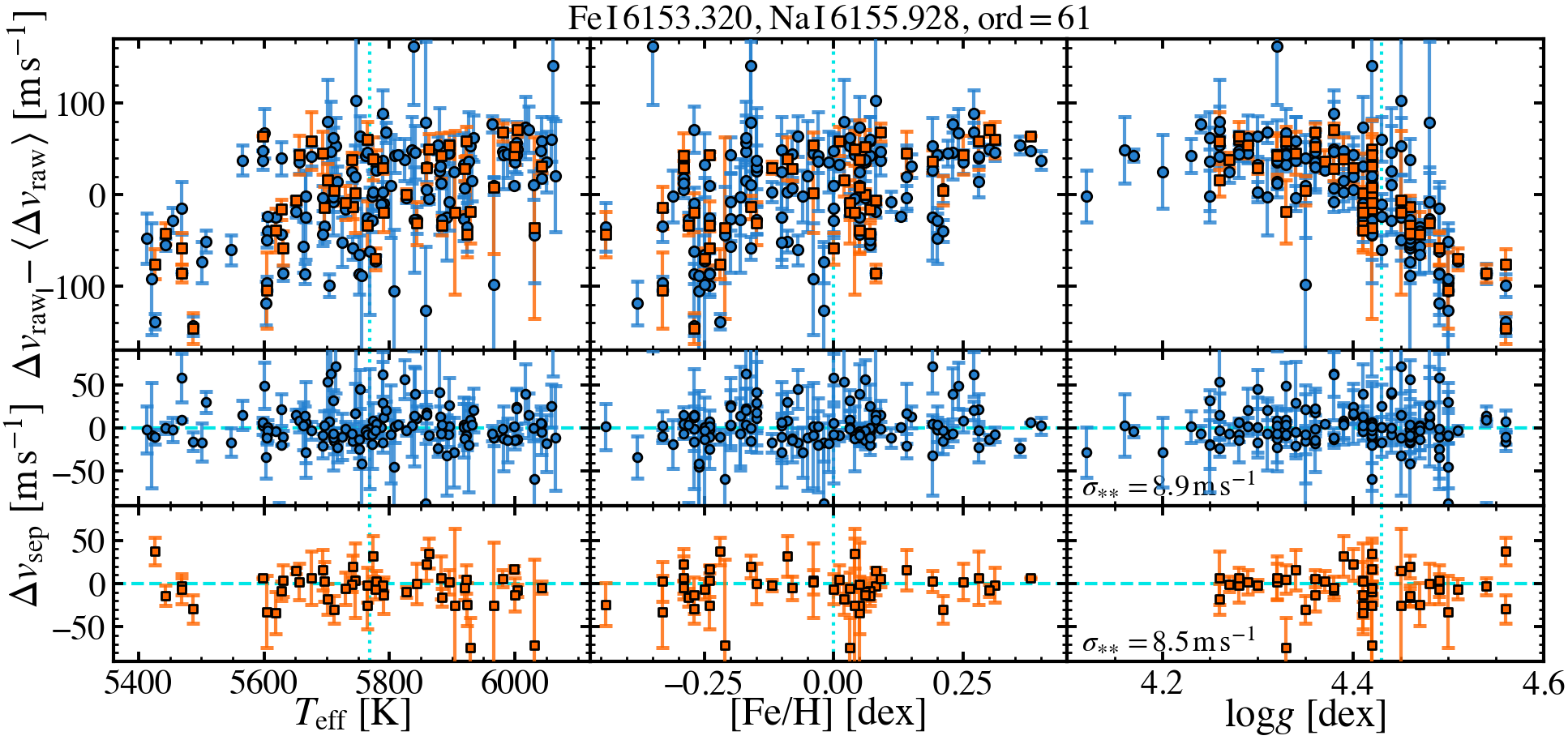}}
\centerline{\includegraphics[width=0.90\columnwidth]{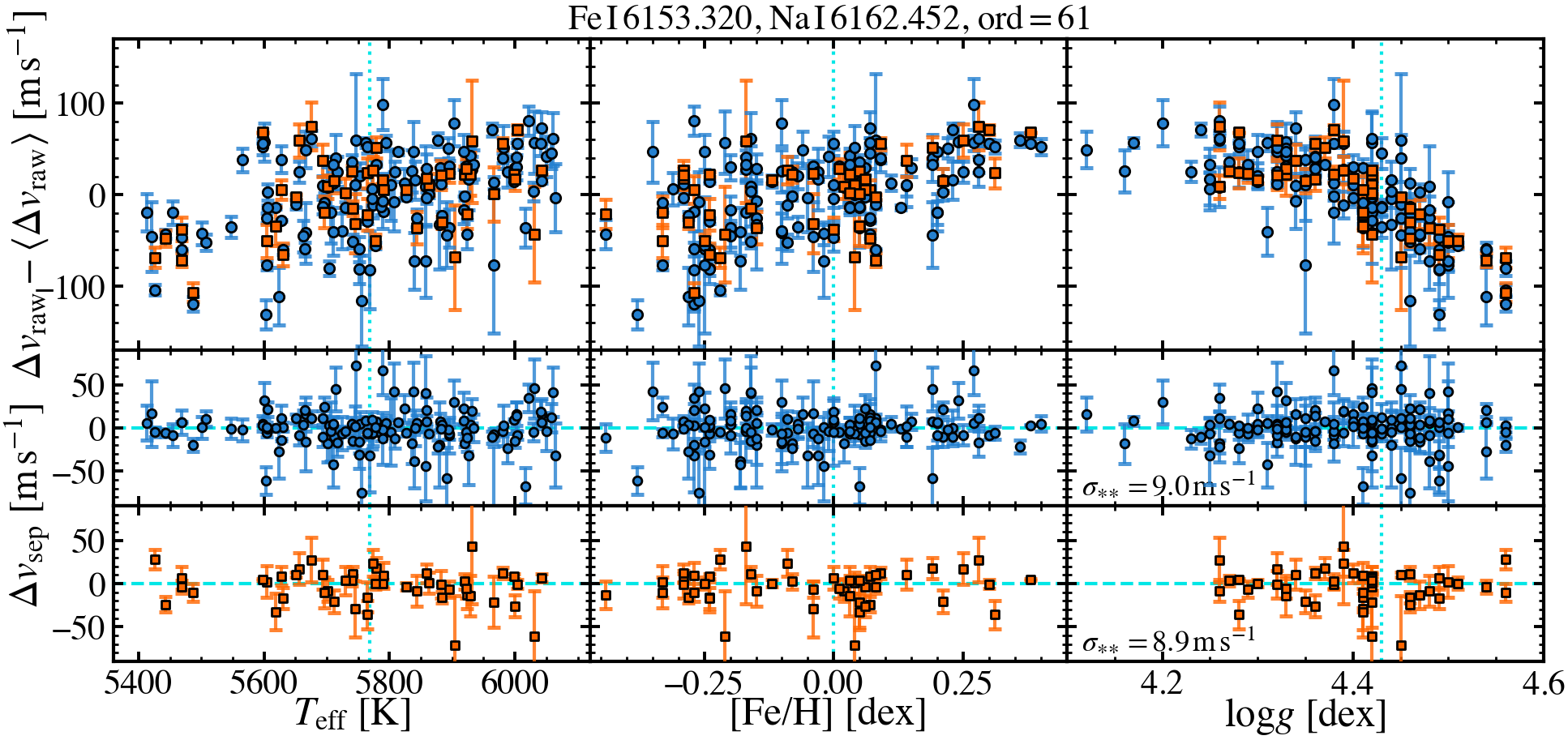}}
\centerline{\includegraphics[width=0.90\columnwidth]{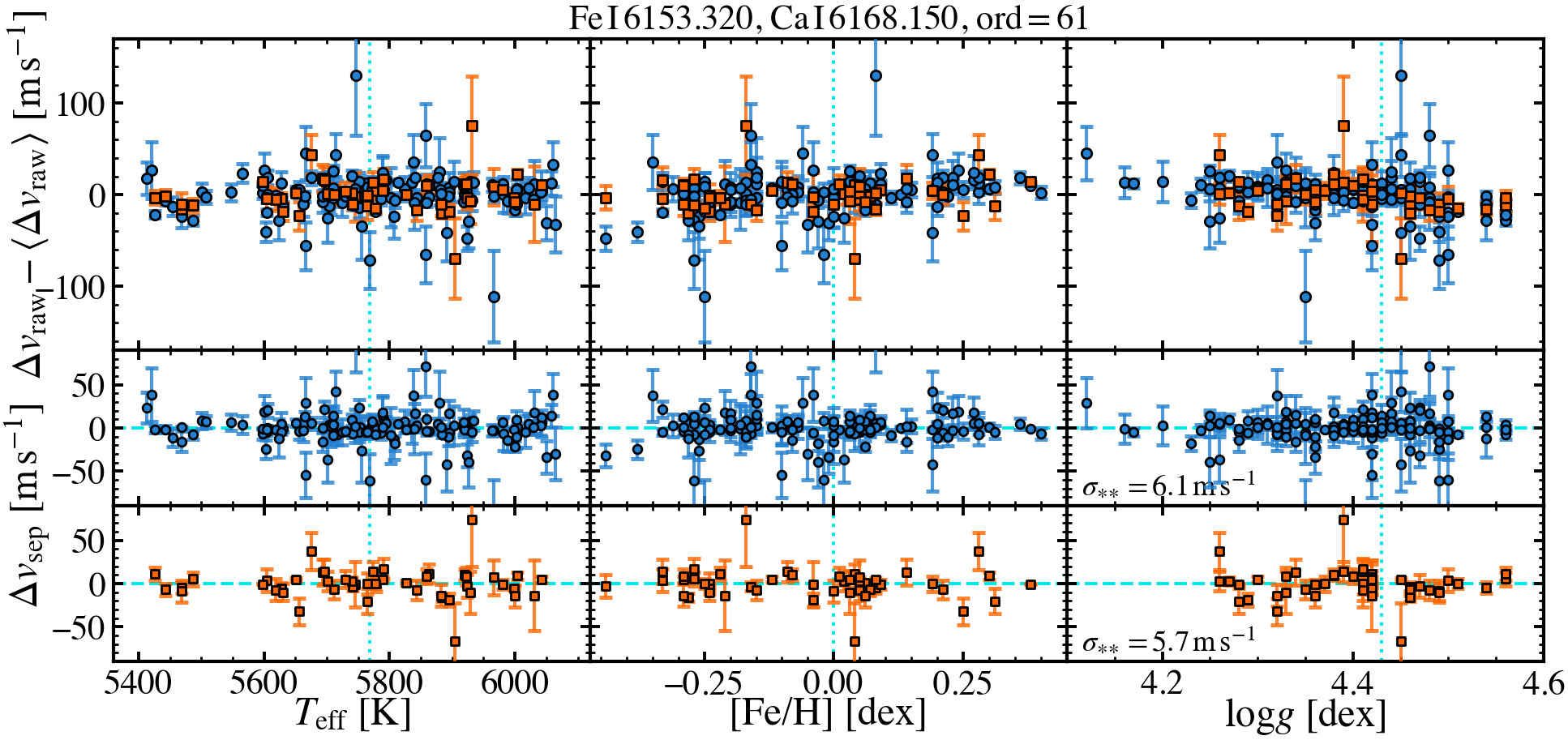}}
}
\caption{
\textbf{Variation in line pair separations.}
}
\end{center}
\end{figure}

\begin{figure}
\begin{center}
\vbox{
\centerline{\includegraphics[width=0.90\columnwidth]{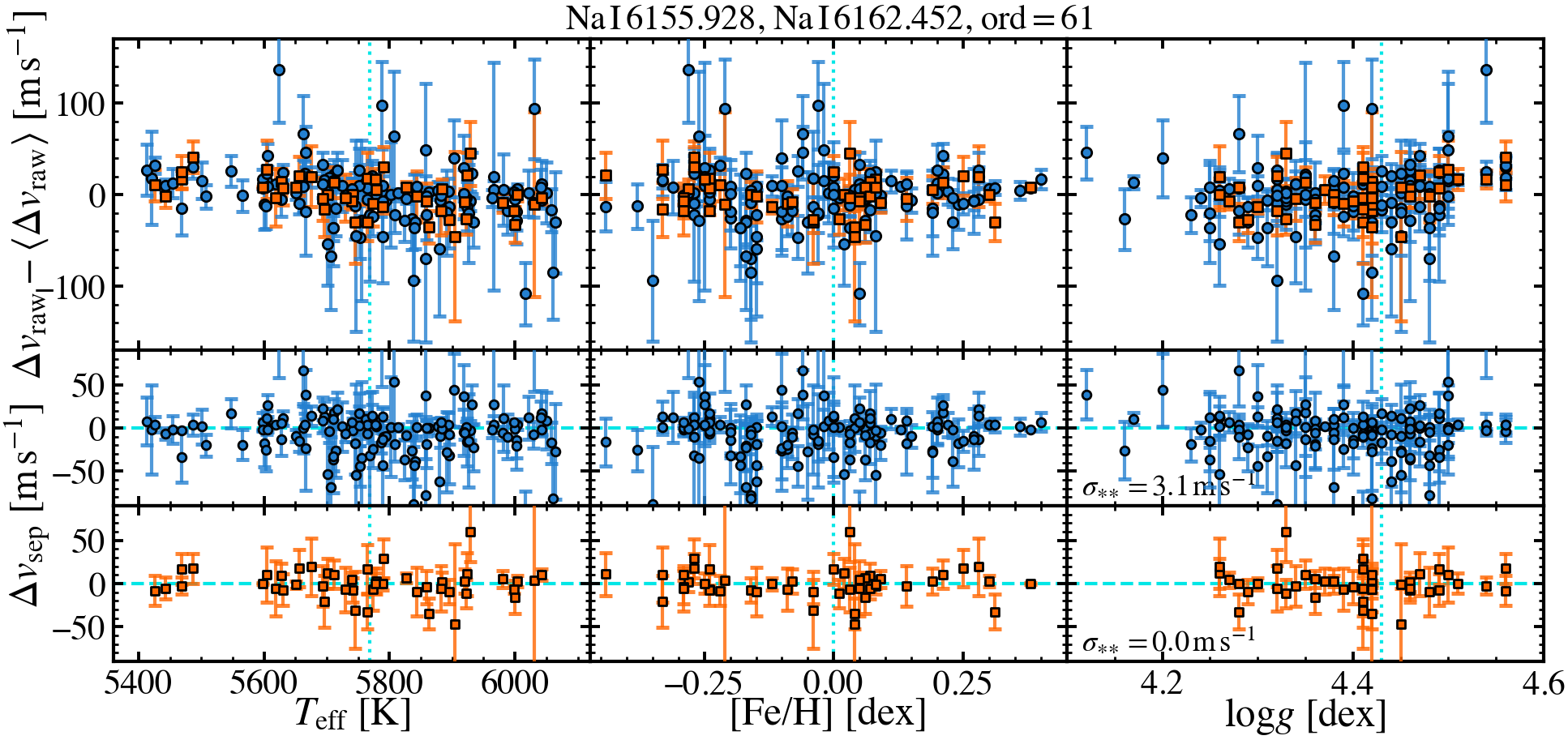}}
\centerline{\includegraphics[width=0.90\columnwidth]{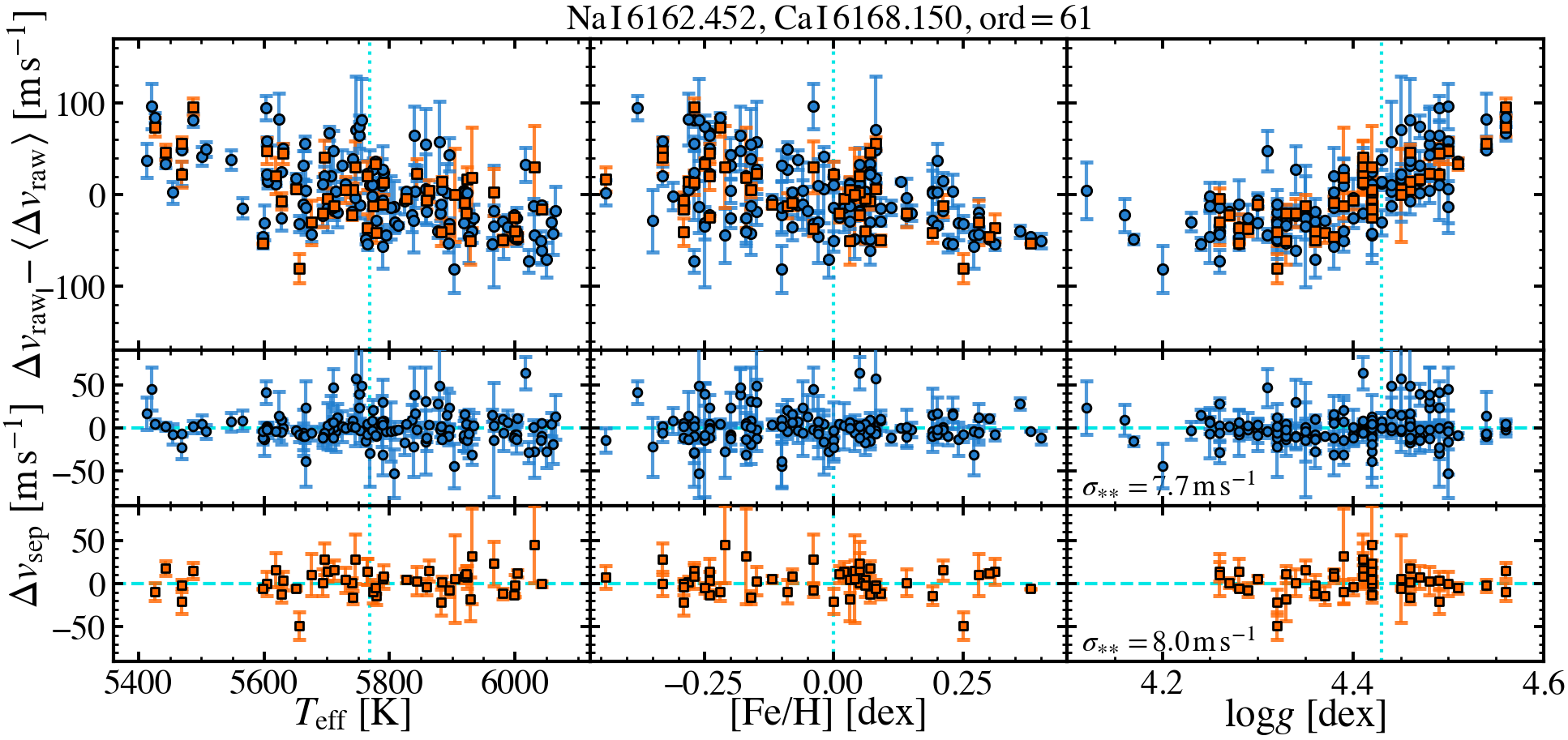}}
\centerline{\includegraphics[width=0.90\columnwidth]{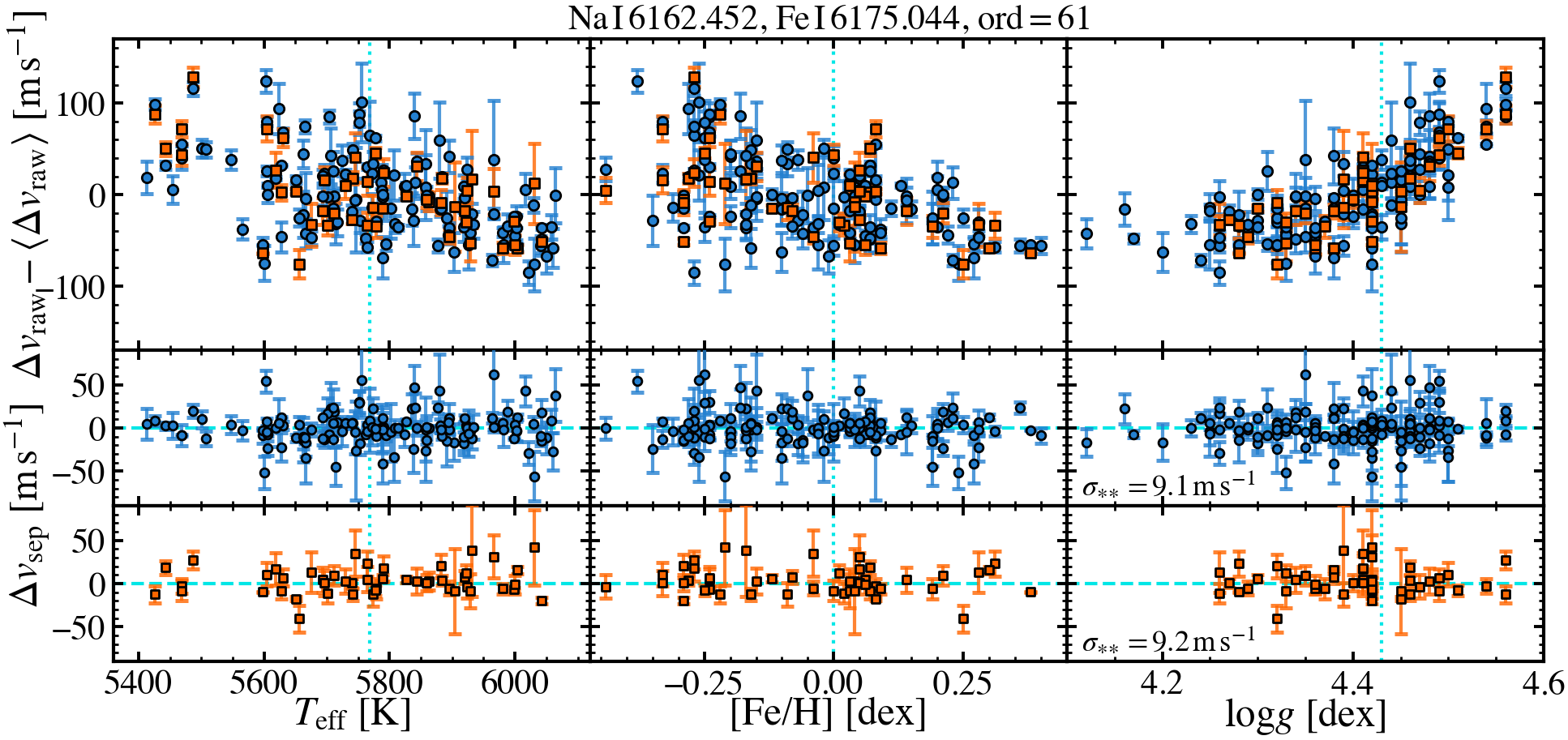}}
}
\caption{
\textbf{Variation in line pair separations.}
}
\end{center}
\end{figure}

\begin{figure}
\begin{center}
\vbox{
\centerline{\includegraphics[width=0.90\columnwidth]{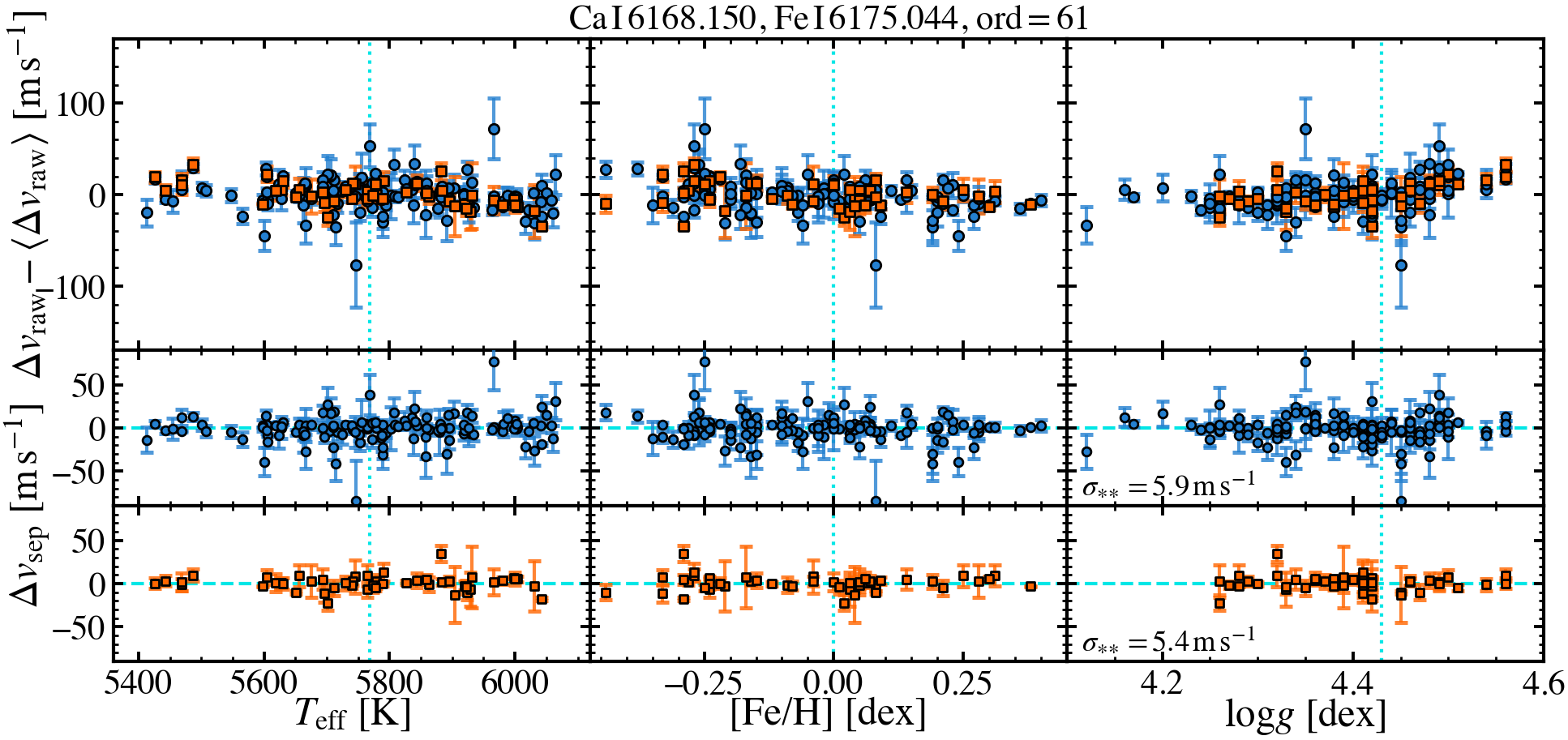}}
\centerline{\includegraphics[width=0.90\columnwidth]{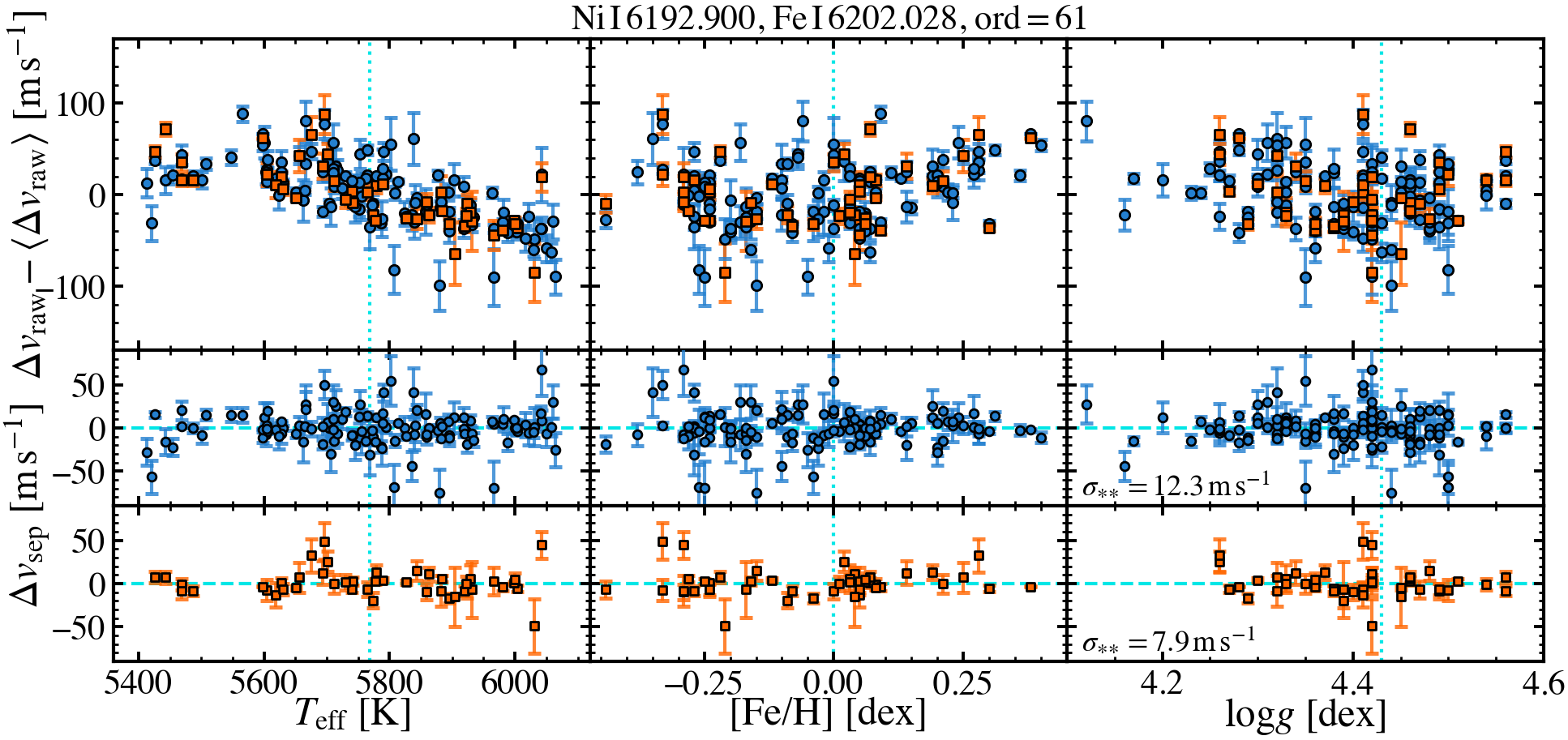}}
\centerline{\includegraphics[width=0.90\columnwidth]{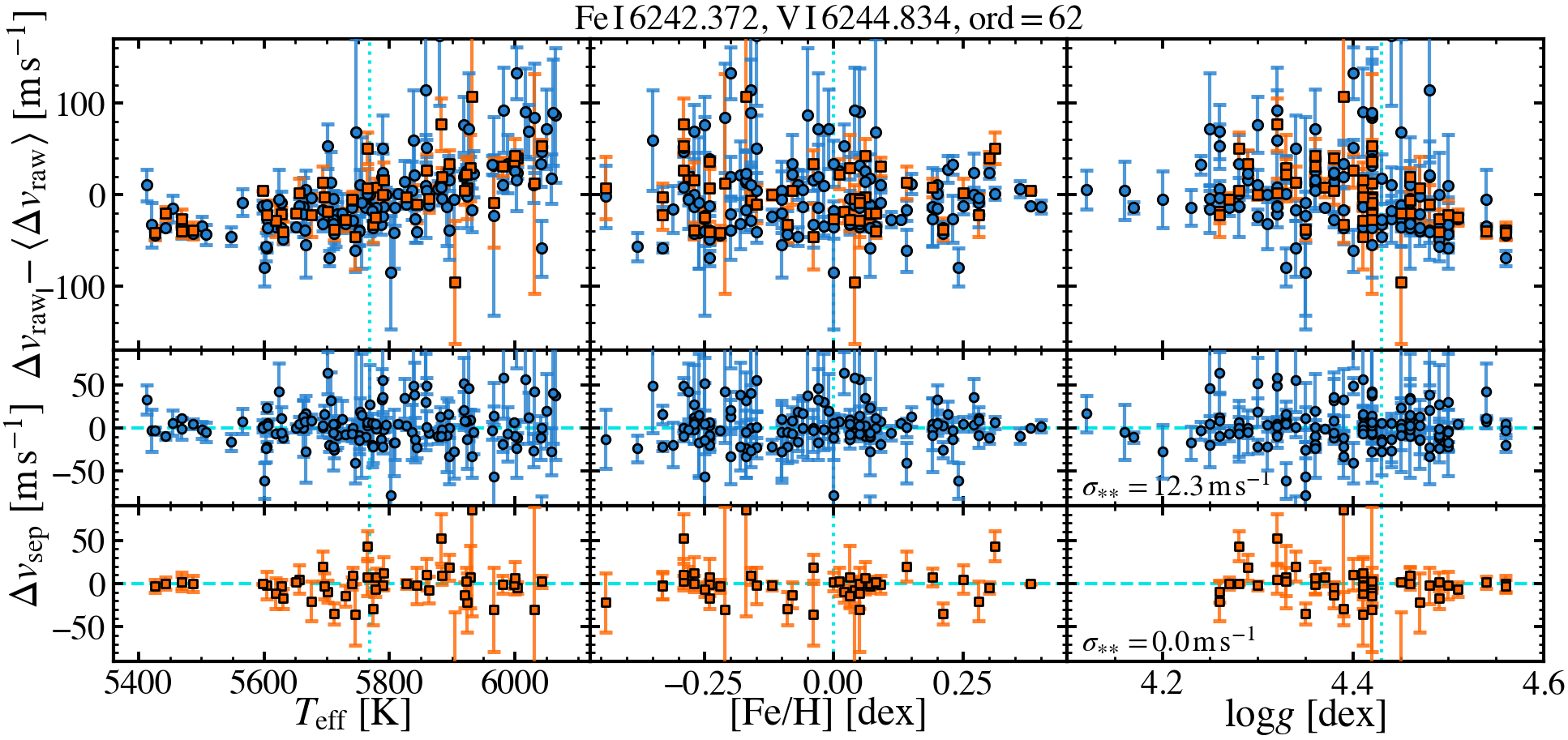}}
}
\caption{
\textbf{Variation in line pair separations.}
}
\end{center}
\end{figure}

A possible alternative approach to estimating $\Delta v^i_{\rm model}$
would be to derive it from simulated stellar line profiles. Numerical
simulations of stellar atmospheres in three dimensions
\cite{Nordlund_1990A&A...228..155N,Asplund_2000A&A...359..729A,Asplund_2009ARA&A..47..481A,Magic_2013A&A...557A..26M}
model the convective motions outlined above and allow individual
spectral line profiles to be synthesised. In principal, a synthetic
profile of an individual line would replace the empirical model,
$\Delta v^i_{\rm model}$, in our approach and could be fitted to the
observed line. However, such models are not accurate enough to avoid
systematic effects in comparing pairs of lines in different (similar)
stars at the $\lsim$5\,m\,s$^{-1}$ level required; the inaccuracies
are up to an order of magnitude larger
\cite{Asplund_2000A&A...359..729A}. In addition, no synthetic profiles
at suitable stellar parameters are available for the 22 lines we use.
We prefer the empirical approach described above because it
incorporates all contributions to variations in the pair separation,
including, for example, from weak blends, and ensures we measure and
account for any remaining star-to-star variance around the model (as
characterised by $\sigma^i_{**}$).

A previous study \cite{Berke_2022b} found that $\Delta v^i_{\rm raw}$
did not vary across exposures of any individual star. Some of the 17
stars had $>$50 high-SNR exposures available, covering more than a
decade of observations, and spanning the change in HARPS optical
fibres in mid-2015. The $\Delta v^i_{\rm raw}$ measurements before the
fibre change were self-consistent, as were those after the fibre
change, but the weighted mean values from before and after the change
disagreed by $\lsim$30\,m\,s$^{-1}$. A similar disagreement was found
in the weighted mean $\Delta v^i_{\rm raw}$ for the two `instances' of
pairs that appeared in two different echelle orders of the
spectrograph \cite{Berke_2022b}. To account for these differences, the
weighted mean $\Delta v^i_{\rm raw}$ was used in the modelling
(averaged across exposures of the same star) where $i$ represents each
instance of a pair before or after the fibre change. That is, for a
single pair, up to 4 different models were used. Fig.\ S2 shows the
result of subtracting the model from the measured
$\Delta v^i_{\rm raw}$ values of a single line pair as a function of
metallicity for all 130 stars in the larger `solar type' sample
\cite{Berke_2022b}, with panels B and C showing the pre- and
post-fibre change results separately. Fig.\ S3 shows the results for
all 17 line pairs used here and all three stellar parameters
($T_{\rm eff}$, [Fe/H] and $\log g$).

For all but 4 of the 17 pairs used to measure $\Delta\alpha/\alpha$,
the statistical uncertainties alone could not explain the scatter
around the model. The additional, star-to-star scatter,
$\sigma^i_{**}$, has been investigated elsewhere \cite{Berke_2022b}
and was found to be mainly astrophysical in origin, although a small
component may be due to instrumental and calibration effects. The
distribution of $\sigma^i_{**}$ for the 17 pairs [and the larger
sample of 229 pairs \cite{Berke_2022b}] peaks around
$\approx$8\,m\,s$^{-1}$, with a range of $\approx$4--15\,m\,s$^{-1}$
for most pairs. There is no evidence for variations in $\sigma^i_{**}$
with $T_{\rm eff}$, [Fe/H] and $\log g$ within the `solar analogue'
range ($\pm$300K, $\pm$0.3\,dex, $\pm$0.4\,dex) \cite{Berke_2022b}, so
we treat $\sigma^i_{**}$ as constant for all solar analogue stars for
a given pair $i$. We expect that the models
\cite{Berke_2022b,BerkeGithub22} should be applicable to all solar
analogue stars.

\subsection*{Deriving \boldmath{$\Delta\alpha/\alpha$} from pair velocity separations}

We calculated the $\Delta\alpha/\alpha$ values from the
$\Delta v^i_{\rm sep}$ and $\sigma^i_{**}$ values measured previously
\cite{Berke_2022a,BerkeGithub22}. The (up to 4) different measurements
of $\Delta v^i_{\rm sep}$ of a single pair, in a single star, were
generally consistent with each other. Therefore, these were averaged
to produce a single $\Delta v^i_{\rm sep}$ value for each pair, $i$,
per star. The weights in this average were taken as the (inverse
square) statistical uncertainties, which are themselves derived from
the weighted mean across all exposures of the star (see above). The
$\sigma^i_{**}$ values were not used in this average, because they are
not meaningful for a single star, and are generally very similar for
the different instances of a pair, both before and after the fibre
change. Including them in the weights has a negligible effect on the
$\Delta\alpha/\alpha$ results. Similarly, to represent the
$\sigma^i_{**}$ value of a pair in a single star, we used the average
of the (up to 4) values for the two instances and fibre-change epochs.

For each star, the $\Delta v^i_{\rm sep}$ values were converted to $\Delta\alpha/\alpha$ using equation (1) applied to a line pair $i$, i.e.\
\renewcommand\theequation{S1}
\begin{equation}
\left(\frac{\Delta\alpha}{\alpha}\right)_i = -\frac{\Delta v^i_{\rm sep}}{c}\frac{\Delta Q}{2}\,,
\end{equation}
where $\Delta Q$ is the difference in $Q$ values of the two lines
comprising the pair (with the same sign convention as the velocity
difference, i.e.\ red minus blue). This has statistical and
star-to-star uncertainties corresponding to those derived for
$\Delta v^i_{\rm sep}$ above, plus an uncertainty in $\Delta Q$
derived from the uncertainties for individual lines
\cite{Dzuba_2022PhRvA.105f2809D}. Within these uncertainties, which
are dominated by those in $\Delta v^i_{\rm sep}$, we found consistency
amongst the $(\Delta\alpha/\alpha)_i$ values for all pairs $i$ for
individual stars. We therefore averaged these together to determine
the $\Delta\alpha/\alpha$ value for each star shown in Fig.\ 2; these
are reported in Table S1. A Monte Carlo simulation was used to derive
this and its statistical and systematic error components: $10^5$
realisations of $(\Delta\alpha/\alpha)_i$ were calculated, drawn from
a Gaussian distribution in $\Delta v^i_{\rm sep}$ with width equal to
the quadrature sum of its statistical and star-to-star uncertainties,
$\sigma^i_{\rm tot}$, and with $\Delta Q$ drawn from a uniform
distribution. We chose the latter because the $Q$ uncertainties are
not derived in a statistical manner, instead representing the range of
plausible theoretical expectations \cite{Dzuba_2022PhRvA.105f2809D}.
Multiple pairs can share the same lines, so uncertainties in their $Q$
values cause correlated errors in $\Delta\alpha/\alpha$. We therefore
derived the $\Delta Q$ distributions from shared distributions of $Q$
for the shared lines. The $\Delta\alpha/\alpha$ value for each star
was calculated as the weighted mean over the realisations, using
$1/(\sigma^i_{\rm tot})^2$ as weights (including uncertainties in
$\Delta Q$ had negligible effect). The statistical and systematic
uncertainties shown in Fig.\ 2 were derived using the same process,
but with the appropriate error components removed, then comparing the
quadrature difference between the resulting distribution of
$\Delta\alpha/\alpha$ and the original one (with all error components
included).

\begin{table}
\renewcommand\thetable{S2}
\caption{\textbf{Stellar parameters and \boldmath{$\Delta\alpha/\alpha$} values.} The first column provides the star names, with the (J2000) right-ascensions and declinations in the second and third columns. The stellar parameters are in the fourth, fifth and sixth columns. The values and statistical uncertainties in $\Delta\alpha/\alpha$ are provided in the seventh and eighth columns, and the final column provides the systematic error derived from the star-to-star scatter values for the 17 line pairs and the $Q$ coefficient uncertainties \cite{Dzuba_2022PhRvA.105f2809D}.}
\label{t:results}
\begin{center}
\begin{tabular}{lcccccccc}\hline
Objects   & RA (J2000)  & Dec.\ (J2000) & $T_{\rm eff}$ & [Fe/H] & $\log g$ & $\Delta\alpha/\alpha$ & $\sigma_{\rm stat}$ & $\sigma_{\rm sys}$ \\
          & (hh:mm:ss)  & (dd:mm:ss)      & (K)          & (dex)  & (dex)    & (ppb)                 & (ppb)             & (ppb)         \\\hline
Sun       & ---         & ---             & 5772         & $ 0.00$  & 4.44     & $ -15.4$ &  26.6 &  36.6 \\
HD~19467  & 03:07:18.58 & $-$13:45:42.4   & 5753         & $-0.07$  & 4.30     & $  70.1$ &  42.5 &  36.9 \\
HD~20782  & 03:20:03.58 & $-$28:51:14.7   & 5773         & $-0.09$  & 4.39     & $ -40.5$ &  30.3 &  33.0 \\
HD~30495  & 04:47:36.29 & $-$16:56:04.0   & 5857         & $-0.02$  & 4.50     & $ -67.2$ & 209.2 &  99.8 \\
HD~45184  & 06:24:43.88 & $-$28:46:48.4   & 5863         & $ 0.04$  & 4.42     & $  25.6$ &   9.6 &  34.1 \\
HD~45289  & 06:24:24.35 & $-$42:50:51.1   & 5710         & $ 0.03$  & 4.25     & $  30.8$ &  25.2 &  36.0 \\
HD~76151  & 08:54:17.95 & $-$05:26:04.0   & 5787         & $ 0.03$  & 4.43     & $-138.3$ &  45.6 &  51.1 \\
HD~78429  & 09:06:38.83 & $-$43:29:31.1   & 5740         & $ 0.05$  & 4.27     & $  10.8$ &  15.2 &  31.3 \\
HD~78660  & 09:09:53.86 & $+$14:27:24.3   & 5788         & $-0.03$  & 4.39     & $-224.4$ & 104.9 &  45.4 \\
HD~138573 & 15:32:43.65 & $+$10:58:05.9   & 5745         & $-0.04$  & 4.41     & $ 158.2$ & 191.7 &  52.6 \\
HD~140538 & 15:44:01.82 & $+$02:30:54.6   & 5693         & $ 0.05$  & 4.46     & $  -9.4$ &  26.8 &  36.9 \\
HD~146233 & 16:15:37.27 & $-$08:22:10.0   & 5826         & $ 0.06$  & 4.42     & $  35.1$ &   7.6 &  30.0 \\
HD~157347 & 17:22:51.29 & $-$02:23:17.4   & 5730         & $ 0.03$  & 4.42     & $  10.8$ &  21.2 &  33.1 \\
HD~171665 & 18:37:12.84 & $-$25:40:16.6   & 5725         & $-0.10$  & 4.46     & $-109.2$ &  56.6 &  40.0 \\
HD~183658 & 19:30:52.72 & $-$06:30:51.9   & 5824         & $ 0.06$  & 4.46     & $  90.3$ &  80.3 &  69.9 \\
HD~220507 & 23:24:42.12 & $-$52:42:06.8   & 5701         & $ 0.02$  & 4.26     & $  61.9$ &  37.1 &  36.3 \\
HD~222582 & 23:41:51.53 & $-$05:59:08.7   & 5802         & $ 0.00$  & 4.35     & $  84.8$ & 134.3 &  42.1 \\\hline
\end{tabular}
\end{center}
\end{table}

The ensemble weighted mean $\Delta\alpha/\alpha$ in equation (3),
averaged over all stars, was calculated with a similar Monte Carlo
approach as above. This is because, in addition to lines (and
therefore $Q$s) being shared between pairs, the star-to-star scatter
terms, $\sigma^i_{**}$, are shared between both pairs and stars. If
these factors are ignored, the weighted mean value of
$\Delta\alpha/\alpha$ in equation (3) is only negligibly affected, but
the systematic error term is underestimated by $\sim$15\%.

\subsection*{Recovery of a large artificial  \boldmath{$\Delta\alpha/\alpha$} signal}

The analysis procedure is completely automated, but the values of some
parameters in the algorithms must be chosen to balance its sensitivity
to $\Delta\alpha/\alpha$ with its ability to reject outliers. An
individual line measurement is rejected if it deviates by more than 3
times its statistical uncertainty from the expected (i.e.\ model)
offset from the laboratory wavelength (see above). If
$\Delta\alpha/\alpha$ is much larger than the uncertainties in one
star, a line in its spectrum with a large $\left|Q\right|$ may be
rejected. For example, the Ni\,{\sc i}\,6192.900 line has $Q=0.18$
\cite{Dzuba_2022PhRvA.105f2809D} so a `large' $\Delta\alpha/\alpha$ of
$+500$\,ppb would shift it bluewards by $\approx$100\,m\,s$^{-1}$,
which is $\sim$4 times larger than the typical measurement uncertainty
from a single HARPS exposure, i.e.\ $\sim$25\,m\,s$^{-1}$. Therefore,
measurements of the line would be rejected in most exposures by our
analysis procedure. However, lines with smaller $\left|Q\right|$ would
be less likely to be rejected. That is, we expect our analysis
procedure to have reduced sensitivity to large differences in $\alpha$
between stars, on average.

To test the response of our analysis procedure to such large
$\Delta\alpha/\alpha$ signals, we shifted the measured wavelengths of
the 22 lines for 8 of the 17 stars by amounts corresponding to
$\Delta\alpha/\alpha = +100$\,ppb (equation 1). We applied the
analysis procedure again, but removed the 8 shifted stars from the
determination of $\Delta v^i_{\rm model}$; this prevents the
artificial shifts from affecting the models and $\sigma^i_{**}$
calculations. Fig.\ S4 shows the change in $\Delta\alpha/\alpha$ for
each star from the original values in Fig.\ 2; Table S3 provides the
numerical values. The weighted mean $\Delta\alpha/\alpha$ for the
shifted stars changed by $+86$\,ppb from the original value, with a
new total uncertainty of 19\,ppb (cf.\ original value of 18\,ppb;
combining statistical and systematic uncertainties in quadrature).
That is, we find that $86 \pm 19$\% of the $+100$\,ppb input signal
was recovered. Nevertheless, Fig.\ S4 also shows that the change in
$\Delta\alpha/\alpha$ for HD~171665 and HD~220507 was $>+$140\,ppb,
while for HD~138573 it was $-175$\,ppb. These deviations from the
expected average response are due to lines and pairs being rejected in
the analysis procedure because of the large $\Delta\alpha/\alpha$
signal. The effect for HD~138573 is particularly large because only
one exposure was available. The unshifted stars show very little
change in $\Delta\alpha/\alpha$; these are due to small changes
in $\Delta v^i_{\rm model}$ because the 8 shifted stars were removed
(out of 130 stars in total) from its determination.

\renewcommand\thefigure{S4}
\begin{figure}
\begin{center}
\centerline{\includegraphics[width=0.67\columnwidth]{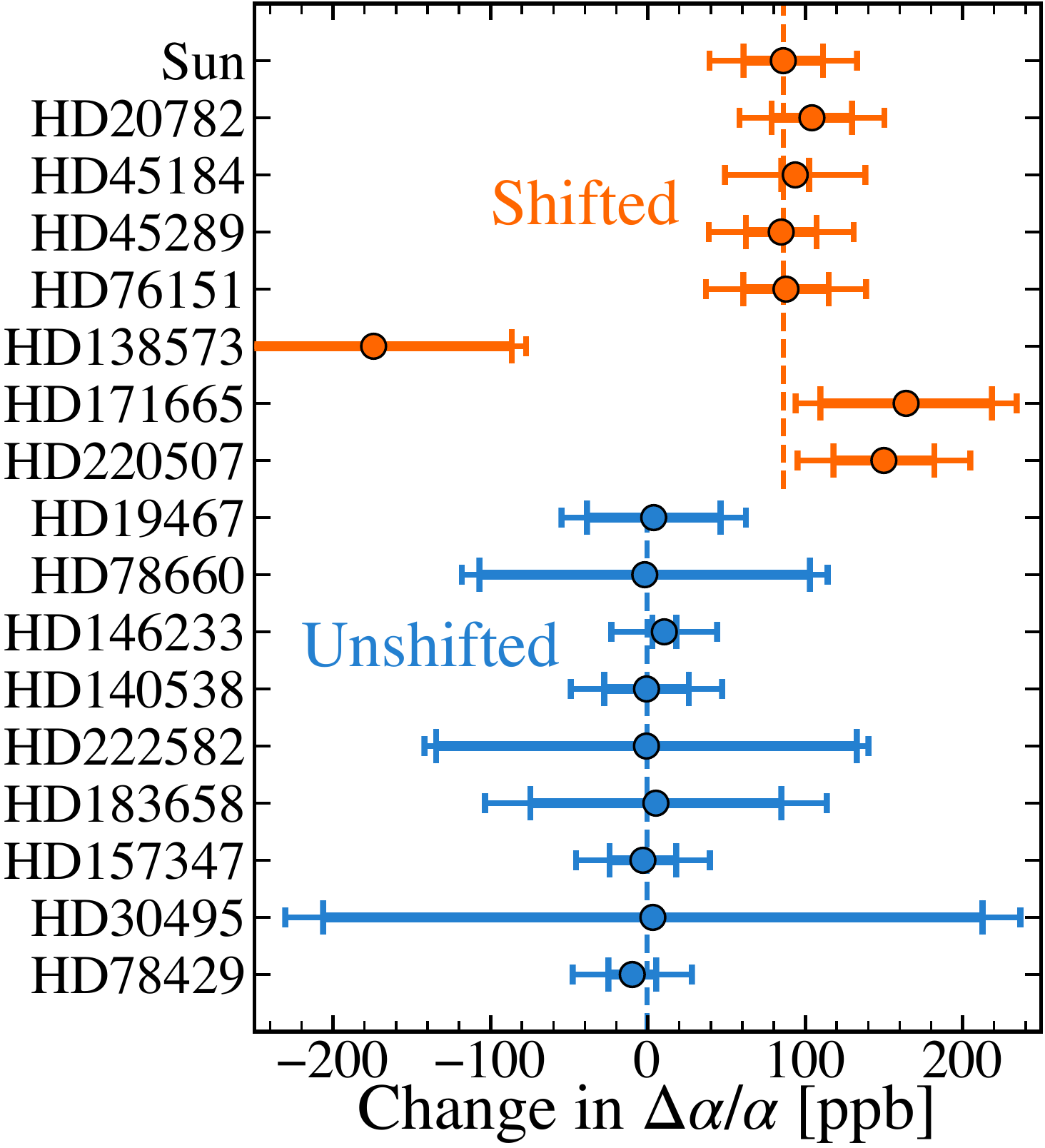}}
\caption{
\textbf{Change in \boldmath{$\Delta\alpha/\alpha$} in response to an artificial signal.} The measured line wavelengths for the 8 shifted stars (orange points) were altered by amounts corresponding to $\Delta\alpha/\alpha = +100$\,ppb, while those for the unshifted stars (blue points) were not altered. The inner (large) and outer (small) error bars have the same meaning as in Fig.\ 2 (i.e.\ statistical and total uncertainties, respectively) but for the new value of $\Delta\alpha/\alpha$ for each star. The dashed vertical lines indicate the change in the weighted means for the shifted (orange) and unshifted (blue) samples.
}
\end{center}
\end{figure}

\begin{table}
\renewcommand\thetable{S3}
\caption{\textbf{Change in \boldmath{$\Delta\alpha/\alpha$} in response to an artificial signal.} The first column provides the star names and the second column states whether the measured line wavelengths for each star were altered using the artificial signal of $\Delta\alpha/\alpha = +100$\,ppb. The change in $\Delta\alpha/\alpha$ from the original results in Fig.\ 2 and Table S2 is shown in the third column. The final two columns show the new $1\sigma$ statistical and systematic uncertainties.}
\label{t:shifts}
\begin{center}
\begin{tabular}{lcccc}\hline
Objects   & Shifted? &  Change in $\Delta\alpha/\alpha$ & $\sigma_{\rm stat}$ & $\sigma_{\rm sys}$ \\
          &          &  (ppb)                           & (ppb)              & (ppb)             \\\hline
Sun       & Yes      & $  85.9 $                        &  25.2              &  39.5 \\
HD~19467  & No       & $   3.2 $                        &  42.5              &  40.3 \\
HD~20782  & Yes      & $ 103.8 $                        &  25.6              &  38.3 \\
HD~30495  & No       & $   4.3 $                        & 209.5              & 103.1 \\
HD~45184  & Yes      & $  93.1 $                        &   8.8              &  43.7 \\
HD~45289  & Yes      & $  84.5 $                        &  22.4              &  40.2 \\
HD~76151  & Yes      & $  87.8 $                        &  27.1              &  43.1 \\
HD~78429  & No       & $ -10.0 $                        &  15.2              &  34.7 \\
HD~78660  & No       & $  -1.8 $                        & 105.0              &  49.8 \\
HD~138573 & Yes      & $-175.1 $                        &  87.7              &  41.1 \\
HD~140538 & No       & $  -0.9 $                        &  26.8              &  39.8 \\
HD~146233 & No       & $  10.0 $                        &   7.6              &  32.7 \\
HD~157347 & No       & $  -3.6 $                        &  21.1              &  36.9 \\
HD~171665 & Yes      & $ 163.1 $                        &  54.5              &  44.2 \\
HD~183658 & No       & $   4.9 $                        &  79.8              &  73.6 \\
HD~220507 & Yes      & $ 149.4 $                        &  32.0              &  44.6 \\
HD~222582 & No       & $  -1.8 $                        & 133.7              &  45.1 \\\hline
\end{tabular}
\end{center}
\end{table}

\end{CJK*}

\end{document}